\documentclass[12pt]{article}

\usepackage{amsmath}
\usepackage{amssymb}
\usepackage{hyperref}
\usepackage{graphicx}

\newcommand{\rem}[1]{}
\newcommand{\remv}[1]{}

\def\amshow{0}

\newcommand\todd{t^{\rm\scriptscriptstyle odd}}
\newcommand\teven{t^{\rm\scriptscriptstyle even}}

\begin{document} 

\begin{titlepage}
\begin{flushright}

\end{flushright}

\begin{center}
{\Large\bf $ $ \\ $ $ \\
Pure spinors in AdS and Lie algebra cohomology
}\\
\bigskip\bigskip\bigskip
{\large Andrei Mikhailov}
\\
\bigskip\bigskip
{\it Instituto de F\'{i}sica Te\'orica, Universidade Estadual Paulista\\
R. Dr. Bento Teobaldo Ferraz 271, 
Bloco II -- Barra Funda\\
CEP:01140-070 -- S\~{a}o Paulo, Brasil\\
}

\vskip 1cm
\end{center}

\begin{abstract}
We show that the BRST cohomology of the massless sector of the Type IIB 
superstring on $AdS_5\times S^5$ can be described as the relative cohomology of an
infinite-dimensional Lie superalgebra. We explain how the vertex operators of 
ghost number 1, which correspond to conserved currents, are described in this
language. We also give some algebraic description of the ghost number 2 
vertices, which appears to be new. We use this algebraic description to 
clarify the structure of the zero mode sector of the ghost number two states 
in flat space, and initiate the study of the vertices of the higher ghost 
number.
\end{abstract}

\end{titlepage}

\tableofcontents
\section{Introduction}
Pure spinor formalism \cite{Berkovits:2000fe} is a generalization of the BRST formalism with the 
ghost fields constrained to satisfy a nonlinear (quadratic) equation:
\begin{equation}\label{IntroPSConstraint}
   \lambda^{\alpha}\Gamma_{\alpha\beta}^m\lambda^{\beta} = 0
\end{equation}
where $\Gamma^m_{\alpha\beta}$ are the Dirac's Gamma-matrices.
A natural question arizes, what kind of nonlinear constraints can ghost fields
satisfy in a physical theory? What if we replace (\ref{IntroPSConstraint}) by an arbitrary set of 
equations: 
\begin{equation}
\lambda^{\alpha}C_{\alpha\beta}^i\lambda^{\beta} = 0\;,\quad
i\in I\quad \mbox{?}
\end{equation}
Of course, this would generally speaking have nothing to do with the string 
theory. But the question is, besides {\em coming from} superstring theory, 
what special properties of $C_{\alpha\beta}^m = \Gamma_{\alpha\beta}^m$ are important for physics? This would 
be useful to know, for example when thinking about possible generalizations of 
the pure spinor formalism. 

It turns out that there is some special property of (\ref{IntroPSConstraint}) which plays an 
important role in the string worldsheet theory. This is the so-called 
Koszulity --- see \cite{Gorodentsev:2006fa} and references therein. The formalism of 
Koszul duality was extensively used in the study of the algebraic properties 
of the supersymmetric Yang-Mills theories in \cite{Movshev:2003ib,Movshev:2004aw}, and in the classification of 
the possible deformations of these theories in \cite{Movshev:2009ba}.

In this paper we will study the BRST cohomology of the massless sector of the 
Type IIB superstring in $AdS_5\times S^5$. We will use the formalism of Koszul 
duality to gain better understanding of the massless BRST cohomology. 

The BRST cohomology counts infinitesimal
deformations of the background $AdS_5\times S^5$, also called ``linearized 
excitations'' or ``gravitational waves''. From the point of view of the string 
worldsheet theory, they are identified with the {\em massless vertex 
operators}. Understanding the properties of these vertex operators is important
already because of their role in the scattering theory. Indeed, the correlation
function of vertex operators is the main ingredient in the string theory 
computation of the S-matrix. 

\paragraph     {Main results} 
\begin{enumerate}
\item We show that the cohomology of the BRST complex of the Type IIB SUGRA on
   $AdS_5\times S^5$ is equivalent to some relative Lie algebra cohomology.
\item We classify the vertex operators of the ghost number 1, which correspond to 
   the densities of the local conserved charges
\item We give a general Lie-algebraic description of the vertex operators of the 
   ghost number $\geq 2$ and use this description to study the properties of 
   the zero momentum states (``discrete states'')
\end{enumerate}

\paragraph     {Previous results for ghost number 1}
The classification of the vertex operators in the ghost number 1 was done, at 
least partially, in the Appendix of our previous paper \cite{Mikhailov:2009rx}; the method which 
we develop here appears more elegant.

\paragraph     {Zero momentum states}
In a typical string theory computation one considers the scattering of 
physical excitations (vertex operators) which depend on the space-time 
coordinates exponentially:
\begin{equation}
   V(x)\simeq e^{ikx}
\end{equation}
But we find it interesting to also consider vertex operators depending on $x$
polynomially. We will call them ``zero momentum vertices'' because their
wavefunction in the momentum space is supported at $k=0$. It turns out that
this ``zero momentum sector'' carries one potentially unpleasant surprize:
there are some well-defined vertex operators which do not correspond to any 
physical states \cite{Bedoya:2010qz,Mikhailov:2012id}. This means that just the requirement of BRST invariance 
alone does not yet provide a complete characterization of the physically 
relevant sigma-models. (But the picture becomes complete if one imposes, in 
addition to the BRST invariance, the condition of the sigma-model being finite 
at the one-loop level.) In this paper we use the Koszul duality to obtain a 
dual description of such unphysical states in terms of fields satisfying 
unusual equations of motion, similar to this one:
\begin{equation}
   \partial_m A_n + \partial_n A_m = 0
\end{equation}
Such equations imply that higher derivatives of $A$ vanish.

\paragraph     {Plan of the paper}
We will start in Sections \ref{sec:Maxwell}, \ref{sec:MaxwellLieCohomology} with the application of Koszul duality to
the ten-dimensional supersymmetric Maxwell theory. In Section \ref{sec:SUGRA} we apply a 
similar method to the study of linearized Type IIB SUGRA in $AdS_5\times S^5$.
We introduce in Section \ref{sec:LieAlgebraOfCovariantDerivatives} some infinite-dimensional super-Lie algebra, and
show in Section \ref{sec:ReductionToIdeal} that the BRST cohomology is equal to the Lie-algebraic
cohomology of some ideal $I$ of this super-algebra. In Section \ref{sec:FlatSpaceLimit} we consider 
the flat space limit and in particular study the zero momentum states. One 
unusual finding is the existence of nontrivial cohomology at the ghost number
three.

\paragraph     {Note added in the revised version}
The approach developed in this paper is useful for clarifying the construction
of {\em integrated} vertex \cite{Chandia:2013kja}.

\section{Pure spinor formulation of the SUSY Maxwell theory}\label{sec:Maxwell}
\subsection{Supersymmetric space-time and basic constraints}
Here we will remind the superspace descirption of the classical 
supersymmetric Maxwell theory in 10 dimensions. The superspace is formed by 
10 bosonic coordinates $x^m$ and 16 fermionic coordinates $\theta^{\alpha}$. This is the
supersymmetric space-time, we will call it $M$:
\begin{equation}
   M = {\bf R}^{10|16}
\end{equation}
The basic 
superfield is the vector potential $A_{\alpha}(x,\theta)$. For every $\alpha\in \{1,\ldots,16\}$,
the corresponding $A_{\alpha}$ is a scalar function:
\begin{equation}
   A_{\alpha}\;:\; M \to {\bf R}
\end{equation}
The equations of motion of the
theory are encoded in the following construction. Let us consider the
``covariant derivatives'':
\begin{equation}\label{CovariantDerivative}
\nabla_{\alpha} = {\partial\over\partial\theta^{\alpha}} + 
\Gamma_{\alpha\beta}^m \theta^{\beta} {\partial\over\partial x^m} + 
A_{\alpha}(x,\theta)
\end{equation}
It turns out \cite{Nilsson:1981bn,Witten:1985nt} that the equations of motion of SUSY Maxwell theory are 
equivalent to the {\em constraint}:
\begin{itemize}
\item There exists a differential operator $\nabla_m = {\partial\over\partial x^m} + A_m(x,\theta)$ such that:
\begin{equation}\label{BasicConstraint}
   \{\nabla_{\alpha},\nabla_{\beta}\} = \Gamma_{\alpha\beta}^m \nabla_m
\end{equation}
\end{itemize}
The nontrivial requirement of the constraint is that the LHS of (\ref{BasicConstraint}) is
proportional to $\Gamma^m_{\alpha\beta}$, because the most general structure would be:
\begin{equation}
\Gamma^m_{\alpha\beta}\nabla_m + \Gamma^{m_1m_2m_3m_4m_5}_{\alpha\beta}X_{m_1m_2m_3m_4m_5}
\end{equation} 
where $X_{m_1\ldots m_5} = X(x,\theta)_{m_1\ldots m_5}$ some function on the superspace. Equivalently,
the constraint (\ref{BasicConstraint}) can be written:
\begin{equation}\label{ContractionWithGammaFiveIsZero}
   \Gamma^{\alpha\beta}_{m_1m_2m_3m_4m_5}\;\{\nabla_{\alpha},\nabla_{\beta}\} = 0
\end{equation}
With the constraint (\ref{ContractionWithGammaFiveIsZero}) satisfied,  we consider (\ref{BasicConstraint}) as the definition of 
$\nabla_m$. The pure spinor interpretation of (\ref{ContractionWithGammaFiveIsZero}) is due to \cite{Howe:1991mf}.

\subsection{Definition of the Lie superalgebra $\cal L$.}\label{sec:LYangMills}
Now let us forget Eq. (\ref{CovariantDerivative}) and consider the Lie superalgebra $\cal L$ generated by 
the letters $\nabla_{\alpha}$ with the relation (\ref{BasicConstraint}). This is an infinite-dimensional Lie
superalgebra. It turns out that some properties of the SUSY Maxwell theory
can be described in terms of this algebra $\cal L$.
In the next Section we will describe an application of the cohomology of ${\cal L}$.

\section{Lie algebra cohomology and solutions of the SUSY Maxwell 
theory}\label{sec:MaxwellLieCohomology}
\subsection{Vacuum solution}
Let us consider the vacuum solution $A_{\alpha}(x,\theta) = 0$. In this case
$\nabla_{\alpha} = \nabla_{\alpha}^{(0)} = {\partial\over\partial\theta^{\alpha}} + 
\Gamma_{\alpha\beta}^m \theta^{\beta} {\partial\over\partial x^m}$. The vacuum solution is invariant under the 
supersymmetry algebra $\bf susy$ generated by the operators $S_{\alpha}$:
\begin{equation}
   S_{\alpha} = {\partial\over\partial\theta^{\alpha}} -
\Gamma_{\alpha\beta}^m \theta^{\beta} {\partial\over\partial x^m}
\end{equation}
We observe that $\{S_{\alpha},\nabla_{\alpha}^{(0)}\} = 0$, and in this sense the vacuum solution is
$\bf susy$-invariant. It turns out that the operators $\nabla^{(0)}$ themselves generate the
same (isomorphic) algebra $\bf susy$ as do $S_{\alpha}$. This can be explained using the 
interpretation of $M$ as the coset space of $\bf susy$. Let us consider the 
abstract algebra $\bf susy$ generated by $\todd_{\alpha}$ and $\teven_m$ with the commutation 
relations:
\begin{equation}
   \{\todd_{\alpha},\todd_{\beta}\} = \Gamma^m_{\alpha\beta} \teven_m
\end{equation}
and other commutators all zero. Let us interpret $x^m$ and $\theta^{\alpha}$ as coordinates on
the group manifold of the corresponding Lie group:
\begin{equation}
   g = \exp(\theta^{\alpha} \todd_{\alpha} + x^m\teven_m)
\end{equation}
Then $\nabla_{\alpha}$ acts as the multiplication by $\todd_{\alpha}$ on the left, and $S_{\alpha}$ as the 
multiplication by $\todd_{\alpha}$ on the right. We can consider the universal enveloping
algebra $U{\bf susy}$ as a representation of $\bf susy$, by the left multiplication.
Then the regular representation can be considered as its dual, which will be
denoted $(U{\bf susy})'$. 

\paragraph     {Relation between $\cal L$ and $\bf susy$.}
There is an ideal $I\subset {\cal L}$ such that the factoralgebra over this ideal 
is $\bf susy$:
\begin{equation}
   {\cal L} / I = {\bf susy}
\end{equation}
The basic constraint (\ref{BasicConstraint}) actually implies the existence of $W^{\alpha}$ such 
that\footnote{A thorough investigation of the consequences of the basic
constraint (\ref{BasicConstraint}) can be found in \cite{Mafra:2009wq}}:
\begin{equation}\label{DefW}
   [\nabla_{\alpha},\nabla_m] = \Gamma^m_{\alpha\beta}W^{\beta}
\end{equation}
This $W^{\alpha}$ is the element of $I$, because if $\nabla_{\alpha}$ were the generators of 
the 10-dimensional supersymmetry algebra, then $W^{\alpha}$ would be zero. 

\subsection{Deformations of solutions and cohomology}\label{sec:DeformationsAndCohomology}
The deformation of the given solution $A_{\alpha}(x,\theta)$ is:
\begin{equation}
A_{\alpha} \mapsto A_{\alpha} + \delta A_{\alpha}
\end{equation}
where $\delta A_{\alpha}$ should satisfy:
\begin{equation}\label{ConstraintOnDeformation}
   \{\nabla_{\alpha}, \delta A_{\beta}\} = \Gamma_{\alpha\beta}^m \delta A_m
\end{equation}
The fact that the LHS is proportional to $\Gamma_{\alpha\beta}^m$ is a nontrivial constraint on 
$\delta A_{\beta}$, and if it is satisfied than (\ref{ConstraintOnDeformation}) becomes the definition of $\delta A_m$.

Let us introduce {\em pure spinors} $\lambda^{\alpha}$ satisfying: 
\begin{equation}
\lambda^{\alpha}\Gamma_{\alpha\beta}^m\lambda^{\beta} = 0
\end{equation}
Using these pure spinors, Eq. (\ref{ConstraintOnDeformation}) can be written:
\begin{align}
   Q v =\;& 0
\\    
\mbox{\tt\small where } Q =\;& \lambda^{\alpha}\nabla_{\alpha} 
\\  
\mbox{\tt\small and } v =\;&  \lambda^{\alpha}\delta A_{\alpha}
\end{align}
Therefore the problem of classifying the infinitesimal deformations of the
vacuum solution is reduced to the computation of the cohomology of $Q$.

\subsection{Koszul duality and its application to deformations}\label{sec:KoszulAndDef}
Let us consider a representation $V$ of the Lie algebra $\bf susy$, and the 
following version of the BRST complex:
\begin{equation}\label{BRSTComplexMaxwellCoeffV}
\ldots  \stackrel{Q_{\rm BRST}}{\longrightarrow} 
   V\otimes_{\bf C}{\cal P}^n
\stackrel{Q_{\rm BRST}}{\longrightarrow} 
   V\otimes_{\bf C}{\cal P}^{n+1}
\stackrel{Q_{\rm BRST}}{\longrightarrow} \ldots
\end{equation}
where ${\cal P}^n$ is the space of polynomial functions of degree $n$ on the pure 
spinors $\lambda^{\alpha}$. A representation $V$ of $\bf susy$ is also a representation of $\cal L$,
because ${\bf susy} = {\cal L}/I$.

Koszul duality\footnote{A nice review can be found in the introductory part of \cite{Gorodentsev:2006fa}; cohomology with coefficients in a representation
was not considered in \cite{Gorodentsev:2006fa}, but it was discussed in
\cite{Movshev:2009ba}} implies that the cohomology of (\ref{BRSTComplexMaxwellCoeffV}) 
coincides with the Lie algebra cohomology of $\cal L$:
\begin{equation}\label{KoszulImpliesBRSTEqualsLie}
   H^n(Q_{\rm BRST}\;;\;V) = H^n({\cal L}\;;\;V)
\end{equation}
Notice that $\bigoplus\limits_{n=0}^{\infty}{\cal P}^n$ is {\em a commutative algebra with quadratic relations}. 
This algebra is Koszul dual to {\em the universal enveloping of a Lie algebra}
$U{\cal L}$.

\paragraph     {Brief review of (\ref{KoszulImpliesBRSTEqualsLie})}
The Koszul duality implies that the following sequence:
\begin{align}\label{KoszulSequence}
   \ldots & \longrightarrow \mbox{Hom}_{\bf C}({\cal P}^2,\;U{\cal L})
   \longrightarrow \mbox{Hom}_{\bf C}({\cal P}^1,\;U{\cal L})
   \longrightarrow U{\cal L} \longrightarrow {\bf C} \longrightarrow 0
\end{align}
is exact, and therefore provides a free resolution of the  $U{\cal L}$-module $\bf C$. 
This fact depends on special properties of the quadratic constraint (\ref{IntroPSConstraint}).

In (\ref{KoszulSequence}) the action of $U{\cal L}$ on $U{\cal L}$ is by the left multiplication, and the 
action of the differential involves the right multiplication by the $\nabla_{\alpha}$:
\begin{equation}
   d \phi(p) = \phi(\lambda^{\alpha}p)\nabla_{\alpha}
\end{equation}
Here on the right hand side we have the product of $\nabla_{\alpha}\in U{\cal L}$ with
$\phi(\lambda^{\alpha}p)\in U{\cal L}$. In other words, for $\phi\in\mbox{Hom}_{\bf C}({\cal P}^n,U{\cal L})$ we have:
\begin{equation}\label{DifferentialInKoszulResolution}
d\phi = \mu^{\rm \tiny right}_{U{\cal L}}(\nabla_{\alpha})\circ\phi\circ\mu_{{\cal P}}(\lambda^{\alpha})
\end{equation}
where  $\mu_{\cal P}(\lambda^{\alpha}):{\cal P}^n\to {\cal P}^{n+1}$ is a multiplication of a polinomial by $\lambda^{\alpha}\in {\cal P}^1$,
and $\mu^{\rm\tiny right}_{U{\cal L}}(\nabla_{\alpha})$ is the right multiplication by $\nabla_{\alpha}$ in $U{\cal L}$. (The composition 
$\phi\circ\mu(\lambda^{\alpha})$ is of the type ${\cal P}^n\to U{\cal L}$; we then multiply by $\nabla_{\alpha}\in U{\cal L}$.)

Since we have a projective resolution of ${\bf C}$, we can now use it to compute
the Lie algebra cohomology of ${\cal L}$ with coefficients in $V$, 
{\it i.e.} $\mbox{Ext}_{U{\cal L}}({\bf C},V)$. It is the cohomology of the following sequence:
\begin{align}
   0 & \longrightarrow \mbox{Hom}_{U{\cal L}}(U{\cal L},V) 
   \longrightarrow 
   \mbox{Hom}_{U{\cal L}}(\mbox{Hom}_{\bf C}({\cal P}^1,U{\cal L}),V)
\longrightarrow\ldots
\label{HomHom}\\   
\ldots & \longrightarrow    
\mbox{Hom}_{U{\cal L}}(\mbox{Hom}_{\bf C}({\cal P}^n,U{\cal L}),V) 
\longrightarrow 
\mbox{Hom}_{U{\cal L}}(\mbox{Hom}_{\bf C}({\cal P}^{n+1},U{\cal L}),V) 
\longrightarrow \ldots
\nonumber
\end{align}
where the differential is induced by (\ref{DifferentialInKoszulResolution}) and acts as follows. 
For $f\in \mbox{Hom}_{U{\cal L}}(\mbox{Hom}_{\bf C}({\cal P}^n,U{\cal L}),V)$, the $df\in\mbox{Hom}_{U{\cal L}}(\mbox{Hom}_{\bf C}({\cal P}^{n+1},U{\cal L}),V)$ 
is evaluated on $\phi\in\mbox{Hom}_{\bf C}({\cal P}^{n+1},U{\cal L})$ as follows:
\begin{equation}
   (df)(\phi: {\cal P}^{n+1} \to U{\cal L}) = 
    f(\mu^{\rm\tiny right}_{U{\cal L}}(\nabla_{\alpha})\circ\phi\circ\mu_{\cal P}(\lambda^{\alpha}))
\end{equation}
There is an isomorphism:
\begin{align}
{\cal P}^n\otimes_{\bf C}V 
\simeq \;& 
  \mbox{Hom}_{U{\cal L}}(\mbox{Hom}_{\bf C}({\cal P}^n,U{\cal L}),V) 
\\    
p\otimes v \mapsto
\;& [\phi\mapsto \phi(p)v]
\end{align}
Here ``$\phi(p)v$'' means the action of $\phi(p)\in U{\cal L}$ on the element $v$ of the 
representation $V$ of $U{\cal L}$. This isomorphism relates (\ref{HomHom}) to (\ref{BRSTComplexMaxwellCoeffV}).

\paragraph     {Special case}
The cohomology problem described in Section \ref{sec:DeformationsAndCohomology} corresponds to the particular
case of $V= (U{\bf susy})'$. As we have just explained, this is equivalent 
to the computation of the Lie algebra cohomology:
\begin{equation}
   H^{\bullet}({\cal L}, (U{\bf susy})')
\end{equation}
Notice that ${\bf susy} = {\cal L}/I$ and therefore $(U{\bf susy})'$ is naturally a 
representation of $\cal L$, by the left multiplication. To calculate this 
cohomology, we notice that the following complex:
\begin{equation}
   \ldots \longrightarrow U{\cal L}\otimes_{\bf C} \Lambda^2 I \longrightarrow
   U{\cal L}\otimes_{\bf C} I \longrightarrow U{\cal L} \longrightarrow 
   U{\bf susy} 
\longrightarrow 0
\end{equation}
is a free resolution of $U{\bf susy}$ as a $U{\cal L}$-module. This means that:
\begin{equation}
   H^n({\cal L}, (U{\bf susy})') = H^n(I,{\bf C})
\end{equation}
More specifically, the ghost number one vertex operator $\lambda^{\alpha}\delta A_{\alpha}$ corresponds to
the first cohomology:
\begin{equation}\label{H1ViaAbelianization}
   H^1(I,{\bf C}) = \left( {I\over [I,I]}\right)'
\end{equation}
This has the following physical interpretation. The space $I\over [I,I]$ can be 
identified with the space of field strengths. Then (\ref{H1ViaAbelianization}) tells us that the 
classical solutions are linear functionals on the space of field strengths. 
Indeed, given a classical solution, we can compute the value of the field 
strenght on this classical solution. Therefore, the space of classical 
solutions is expected to be dual to the space of field strengths, as we indeed
observe in (\ref{H1ViaAbelianization}). 

\paragraph     {Explicit description of $I\over [I,I]$}
Elements $W^{\alpha}$ of $I$ were introduced in Eq. (\ref{DefW}).  Consider
the projection of $W^{\alpha}$ to $I/[I,I]$, {\it i.e.} $W^{\alpha} \mbox{ mod } [I,I]$. We conjecture 
that all the other elements of $I/[I,I]$ can be obtained from $W^{\alpha}$ by commuting
with $\nabla_{\alpha}$, {\it i.e.} acting with $\bf susy$. This means that all the gauge 
invariant operators at the linearized level are $W^{\alpha}$ and its derivatives.

\section{Type IIB SUGRA in $AdS_5\times S^5$}\label{sec:SUGRA}
\paragraph     {Note in the revised version}
The constructions of this paragraph can be illustrated by explicit examples
of vertex operator, corresponding  to the $\beta$-deformation
These  examples are constructed in \cite{Chandia:2013kja}.

\subsection{BRST complex}\label{sec:BRSTComplexTypeIIB}
The BRST complex of Type IIB SUGRA in $AdS_5\times S^5$ \cite{Berkovits:2001ue,Berkovits:2000yr,Berkovits:2004xu} is based on the coset 
space $G/G_0$ where $G$ is the Lie supergroup corresponding to the Lie 
superalgebra ${\bf g} = {\bf psu}(2,2|4)$ and $G_0$ is the subgroup corresponding to 
${\bf g}_{\bar{0}} = so(1,4)\oplus so(5)$. A ${\bf Z}_4$-grading of $\bf g$ plays an important role. The 
generators of $\bf g$ are denoted:
\begin{align}
t^3_{\alpha} & \mbox{ \tt\small of degree 3},\; \alpha\in \{1,\ldots,16\}   
\nonumber \\   
t^1_{\dot{\alpha}} & \mbox{ \tt\small of degree 1},\; \dot{\alpha}\in\{1,\ldots,16\} 
\nonumber \\  
t^2_n & \mbox{ \tt\small of degree 2}, \; n \in \{0,\ldots,9\}
\label{BasisOfPSU}
\\     
t^0_{[mn]} & \mbox{ \tt\small of degree 0}
\nonumber
\end{align}
The subalgebra ${\bf g}_{\bar{0}}$ is generated by $t^0_{[mn]}$, ${\bf g}_{\bar{3}}$ by $t^3_{\alpha}$, ${\bf g}_{\bar{1}}$ by $t^1_{\dot{\alpha}}$, and ${\bf g}_{\bar{2}}$ by $t^2_m$. 
The index $[mn]$ of $t^0_{[mn]}$ runs over a union of two sets: the set of choices of 
2 different elements $m,n$ from $\{0,\ldots 4\}$, and the set of choices of 2 
different elements $m,n$ from $\{5,\ldots,9\}$. This corresponds to the split of ${\bf g}_{\bar{0}}$
into the direct sum of $so(1,4)$ and  $so(5)$. Both $t_{\alpha}^3$ and $t_{\dot{\alpha}}^1$ transform as 
spinors of both $so(1,4)$ and $so(5)$ under the adjoint action of ${\bf g}_{\bar{0}}$, and $t^2_m$ 
transform as vectors.

The BRST complex computing supergravity excitations on the background
$AdS_5\times S^5$ is:
\begin{align}\label{StandardBRSTComplex}
\ldots  \stackrel{Q_{\rm BRST}}{\longrightarrow} \mbox{Hom}_{{\bf g}_{\bar{0}}}\left(
   U{\bf g}\;,\;{\cal P}^n
\right) \stackrel{Q_{\rm BRST}}{\longrightarrow} \mbox{Hom}_{{\bf g}_{\bar{0}}}\left(
   U{\bf g}\;,\;{\cal P}^{n+1}
\right) \stackrel{Q_{\rm BRST}}{\longrightarrow} \ldots
\end{align}
where ${\cal P}^n$ is the space of polynomials functions of the order $n$ of two
independent pure spinors $\lambda_L$ and $\lambda_R$: 
\begin{equation}
   \lambda_L^{\alpha}f_{\alpha\beta}{}^m\lambda_L^{\beta} = 0\;,\quad
   \lambda_R^{\dot{\alpha}}f_{\dot{\alpha}\dot{\beta}}{}^m\lambda_R^{\dot{\beta}} = 0
\quad \mbox{ for } m\in \{0,\ldots,9\}
\end{equation}
where $f_{\bullet\bullet}{}^{\bullet}$ are the structure constants of ${\bf g}$, and $Q_{\rm BRST}$ is given by:
\begin{align}
Q_{\rm BRST} = \;& Q_{\rm BRST}^L + Q_{\rm BRST}^R
\\    
\mbox{\tt\small where } Q_{\rm BRST}^L =\;& 
\lambda_L^{\alpha}L(t^3_{\alpha}) 
\\   
\mbox{\tt\small and } Q_{\rm BRST}^R =\;&
\lambda_R^{\dot{\alpha}}L(t^1_{\dot{\alpha}})
\end{align}
Here $L(t)$ is the left multiplication by $t$. We will use the notation ${\cal P}^{p,q}$ for 
the space of polynomials of the order $p$ in $\lambda_L$ and $q$ in $\lambda_R$. Therefore 
${\cal P}^n =\bigoplus_{p+q=n}{\cal P}^{p,q}$.

More generally, we can consider the cohomology with coefficients in an
arbitrary representation $V$ of ${\bf g}$:
\begin{equation}\label{BRSTComplexGeneralRepresentation}
\ldots  \stackrel{Q_{\rm BRST}}{\longrightarrow} 
   V\otimes_{{\bf g}_{\bar{0}}}{\cal P}^n
\stackrel{Q_{\rm BRST}}{\longrightarrow} 
   V\otimes_{{\bf g}_{\bar{0}}}{\cal P}^{n+1}
\stackrel{Q_{\rm BRST}}{\longrightarrow} \ldots
\end{equation}
The cohomology of this complex\footnote{Frobenius reciprocity implies a relation between (\ref{StandardBRSTComplex}) and (\ref{BRSTComplexGeneralRepresentation}), see \cite{Mikhailov:2009rx}.} will be denoted $H^n(Q_{\rm BRST}\;;\;V)$. 
With this notation, the cohomology of the ``standard'' BRST complex (\ref{StandardBRSTComplex}) is
$H^n(Q_{\rm BRST}\;;\; (U{\bf g})')$. These complexes were studied in \cite{Berkovits:2000yr,Mikhailov:2009rx,Mikhailov:2011af}.

It is useful to consider a {\em filtration} $F^p$ on the space of vertex 
operators, corresponding to the powers of $\lambda_R$. We will consider an element of 
$\mbox{Hom}_{{\bf g}_{\bar{0}}}(U{\bf g}\;,\;{\cal P}^n)$ to be of the order $p$ if it goes like $O(\lambda_R^p)$ when $\lambda_R \to 0$. 
The space of such operators will be donoted $F^p \mbox{Hom}_{{\bf g}_{\bar{0}}}(U{\bf g}\;,\;{\cal P}^n)$. This is a
decreasing filtration, {\it i.e.} $\ldots \supset F^p\supset F^{p+1} \supset F^{p+2}\supset\ldots$ This allows
us to calculate the cohomology of $Q_{\rm BRST}$ using some approximation scheme,
starting from the cohomology of $Q^L_{\rm BRST}$ and considering $Q^R_{\rm BRST}$ as a small 
correction. The first approximation is:
\begin{equation}
   E_2^{p,q} = H^p(Q^R_{\rm BRST}\;;\; H^q(Q^L_{\rm BRST}\;;\;V))
\end{equation}

\subsection{Lie algebra formed by the covariant derivatives}\label{sec:LieAlgebraOfCovariantDerivatives}
Now we will introduce some infinite-dimensional Lie algebra, which we will
use later to study the cohomology of the complexes (\ref{StandardBRSTComplex}) and (\ref{BRSTComplexGeneralRepresentation}).

\paragraph     {Definition of the Lie algebra ${\cal L}^{\rm tot}$.}
We will consider the infinite-dimensional super Lie algebra generated by the 
following letters:
\begin{equation}
   \nabla^L_{\alpha} \;,\; \nabla^R_{\dot{\alpha}} \;,\; t^0_{[mn]}
\end{equation}
where the indices $\alpha$, $\dot{\alpha}$ and $[mn]$ run over the same sets as in (\ref{BasisOfPSU}), 
and with the following relations:
\begin{align}
\{\nabla_{\alpha}^L\;,\;\nabla_{\beta}^L\} = \;& f_{\alpha\beta}{}^m\nabla^L_m
\label{DefNablaML}\\     
\{\nabla_{\dot{\alpha}}^R\;,\;\nabla_{\dot{\beta}}^R\} = \;& 
f_{\dot{\alpha}\dot{\beta}}{}^m\nabla^R_m
\label{DefNablaMR}\\     
\{\nabla_{\alpha}^L\;,\;\nabla_{\dot{\beta}}^R\} = \;&
f_{\alpha\dot{\beta}}{}^{[mn]} t^0_{[mn]}
\label{CollapsToT0}\\   
[t^0_{[mn]}\;,\;\nabla_{\alpha}^L] =\;&
f_{[mn]\alpha}{}^{\beta}\nabla_{\beta}^L
\label{ActionOfT0OnLeft}\\   
[t^0_{[mn]}\;,\;\nabla_{\dot{\alpha}}^R] =\;&
f_{[mn]\dot{\alpha}}{}^{\dot{\beta}}\nabla_{\dot{\beta}}^R
\label{ActionOfT0OnRight}\\    
[t^0_{[kl]}\;,\;t^0_{[mn]}] =\;& f_{[kl][mn]}{}^{[pq]}t^0_{[pq]}
\end{align}
where Eqs. (\ref{DefNablaML}) and (\ref{DefNablaMR}) are the definitions of $\nabla^L_m$ and $\nabla^R_m$. The 
coefficients $f_{\bullet\bullet}{}^{\bullet}$ are the structure constants of $psu(2,2|4)$ in the basis
(\ref{BasisOfPSU}). We will introduce the following notation for this Lie algebra:
\begin{equation}
   {\cal L}^{\rm tot} = {\cal L}^L + {\cal L}^R + {\bf g}_{\bar{0}}
\end{equation}
where the sum is as linear spaces. More details are in \cite{Mikhailov:2013vja}.

\paragraph     {Grading.}
We will introduce on  ${\cal L}^{\rm tot}$ a ${\bf Z}$-grading as follows:
\begin{align}
\mbox{deg}(\nabla_{\alpha}^L) = \;& 1
\nonumber \\   
\mbox{deg}(\nabla_{\dot{\alpha}}^R) = \;& -1
\label{ZGrading}
\end{align}

\paragraph     {Definition of the ideal $I\subset {\cal L}^{\rm tot}$.}  
There is an ideal $I\subset {\cal L}^{\rm tot}$ such that ${\cal L}^{\rm tot}/I = {\bf g}$. The structure of $\bf g$ is 
explained in Eq. (\ref{BasisOfPSU}). Modulo $I$ the generators $t^0_{[mn]}$ become the generators 
$t^0_{[mn]}$ of ${\bf g}_{\bar{0}}\subset {\bf g}$, $\nabla^L_{\alpha}$ becomes $t^3_{\alpha}$, $\nabla^R_{\dot{\alpha}}$ becomes $t^1_{\dot{\alpha}}$, and both $\nabla^L_m$ and $\nabla^R_m$ 
become $t^2_m$. The ideal $I$ is not invariant under the $U(1)$ which defines the
${\bf Z}$-grading (\ref{ZGrading}), but only under ${\bf Z}_4\subset U(1)$.

\subsection{Lie algebra cohomology}\label{sec:LieAlgebraCohomology}
Let us consider the relative Lie algebra cohomology\footnote{For introduction into the Lie algebra cohomology, see \cite{Knapp,FeiginFuchs}}:
\begin{equation}\label{RelativeLieCohomology}
   H^{\bullet}\left( {\cal L}^{\rm tot}\;;\; {\bf g}_{\bar{0}}\;;\; V\right)
\end{equation}
We claim that this cohomology coincides with the BRST cohomology:
\begin{equation}\label{RelativeCohomologyCoincidesWithBRST}
     H^{\bullet}\left( {\cal L}^{\rm tot}\;;\; {\bf g}_{\bar{0}}\;;\; V\right)
= H(Q_{\rm BRST}\;;\;V)
\end{equation}
We will prove a stronger statement. Let us introduce a decreasing filtration
of the Lie algebra cochain complex in the following way. We say that a cochain 
$c$ belongs to $F^pC^q\left( {\cal L}^{\rm tot}\;;\; {\bf g}_{\bar{0}}\;;\; V\right)$ if $c(\xi_1,\ldots,\xi_q)$ is zero whenever there are 
less than $p$ letters $\nabla_{\dot{\alpha}}^R$ among $\xi_1,\ldots,\xi_q$. For example, for $c\in F^3C^2$ should
be true that $c(\nabla^R_{\dot{\alpha}},\nabla^R_{\dot{\beta}}) = 0$, but $c(\nabla^R_m,\nabla^R_{\dot{\beta}})$ does not have to be zero 
(because $\nabla^R_m$ is defined in (\ref{DefNablaMR}) as the commutator of two $\nabla_{\dot{\alpha}}^R$, {\it i.e.} has
degree 2).

In other words, the ghost dual to $\nabla_{\dot{\alpha}}^R$ is considered ``small of the order $\varepsilon$''; 
the ghost dual to $\nabla_m^R$ is considered ``small of the order $\varepsilon^2$'', {\it etc.} But 
all the ``left'' ghosts are of the order 1. The $F^pC$ consists of cochains 
which are of the order $\varepsilon^p$ and higher. 

Similarly, the BRST complex has a filtration by the powers of $\lambda_R$.

We will construct a {\em filtered} quasi-isomorphism between the relative Lie 
algebra complex $C^{\bullet}({\cal L}^{\rm tot}\;;\; {\bf g}_{\bar{0}};\;;V)$ and the BRST complex. A filtered 
quasi-isomorphism of two filtered complexes $C_1^{\bullet}$ and $C_2^{\bullet}$ is a map of complexes 
which is a quasi-isomorphism ${\bf gr}^pC_1^{\bullet} \to {\bf gr}^pC_2^{\bullet}$ for every $p$. A filtered 
quasi-isomorphism is a quasi-isomorphism of complexes in the usual sense, if 
one forgets the grading \cite[Lemma 05S3]{stacks-project}. This can be understood from the point of view of 
spectral sequences; filtered quasi-isomorphism becomes an isomorphism at $E_1^{\bullet,\bullet}$.

In particular, it follows  that the relative Lie algebra cohomology (\ref{RelativeLieCohomology}) 
coincides with the BRST cohomology (\ref{BRSTComplexGeneralRepresentation}). 

\paragraph     {Construction of filtered quasi-isomorphism.}
Let $C^{\bullet}({\cal L}^{\rm tot}\;;\; {\bf g}_{\bar{0}}\;;\;V)$ denote the space of  cochains in the relative Lie 
algebra cohomology complex (\ref{RelativeLieCohomology}). Let us introduce the operation of 
restriction from the space of relative cochains to the BRST complex:
\begin{equation}
   R\;:\;\;C^{\bullet}({\cal L}^{\rm tot}\;;\; {\bf g}_{\bar{0}}\;;\;V)
\longrightarrow V\otimes \mbox{Fun}(\lambda_L,\lambda_R)
\end{equation}
which is defined as follows. Given the cochain $c\in C^q({\cal L}^{\rm tot}\;;\; {\bf g}_{\bar{0}}\;;\;V)$, we have
to define $Rc\in V\otimes \mbox{Fun}(\lambda_L,\lambda_R)$. By definition $c$ is a polylinear function of
$q$ elements of ${\cal L}$:
\begin{equation}
   \xi_1\wedge\xi_2\wedge\cdots\wedge\xi_q \mapsto 
c(\xi_1\wedge\xi_2\wedge\ldots\wedge\xi_q)
\end{equation}
Elements of the linear space ${\cal L}^{\rm tot}/{\bf g}_{\bar{0}}$ are, by definition in Section \ref{sec:LieAlgebraOfCovariantDerivatives}, 
nested commutators of $\nabla^L$s plus nested commutators of $\nabla^R$s. We define $Rc$ as 
the following function of $\lambda_L$ and $\lambda_R$:
\begin{align}
Rc(\lambda_L,\lambda_R) =\;& c\left(
   \left(
      \lambda^{\alpha}_L\nabla^L_{\alpha} + 
      \lambda^{\dot{\alpha}}_R\nabla^R_{\dot{\alpha}}
   \right)^{\otimes q}
\right)
\\    
& \mbox{\tt\small for } c \in 
C^{q}({\cal L}^{\rm tot}\;;\; {\bf g}_{\bar{0}}\;;\;V)
\nonumber
\end{align}
We used the following notation: $\xi^{\otimes q}$ means $\underbrace{\xi \otimes \xi \otimes \cdots \otimes \xi}_{q \text{ times}}$. We observe:
\begin{equation}
   RQ_{\rm Lie} = Q_{\rm BRST} R
\end{equation}

\paragraph     {Lemma:} $R$ is a filtered quasi-isomorphism.

\noindent
To prove this, we consider the action of $Q_{\rm Lie}$ on the following space:
\begin{align}
{\bf gr}^pC^{p+q}({\cal L}^{\rm tot}\;;\; {\bf g}_{\bar{0}}\;;\;V)
= \;&
{F^pC^{p+q}({\cal L}^{\rm tot}\;;\; {\bf g}_{\bar{0}}\;;\;V)
\over 
F^{p+1}C^{p+q}({\cal L}^{\rm tot}\;;\; {\bf g}_{\bar{0}}\;;\;V)} =
\nonumber \\   
= \;& \bigoplus^p_{r=0}
C^{q+r}({\cal L}^L;V)\otimes_{{\bf g}_0} 
{\bf gr}^p C^{p-r}({\cal L}^R;{\bf C})
\end{align}
We observe that:  
\begin{enumerate}
\item The action of $Q_{\rm Lie}$ on ${\bf gr}^pC^{p+q}({\cal L}^{\rm tot}\;;\; {\bf g}_{\bar{0}}\;;\;V)$ coincides with the action of 
   the operator $Q_{\rm Lie}^{[H^{\bullet}({\cal L}^L,V)]} + Q_{\rm Lie}^{[H^{\bullet}({\cal L}^R,{\bf C})]}$ on 
   $\bigoplus_{r=0}^p C^{q+r}({\cal L}^L;V)\otimes_{{\bf g}_0} {\bf gr}^p C^{p-r}({\cal L}^R;{\bf C})$
\item The restriction map $R$ is only nonzero on the $r=0$ term. It intertwines 
   this complex with the left BRST complex, which has the BRST 
   operator $Q_L = \lambda_L^{\alpha} t^3_{\alpha}$. In other words, it is a morphism of complexes:
\begin{equation}
   {\bf gr}^pC^{p+q}({\cal L}^{\rm tot}\;;\; {\bf g}_{\bar{0}}\;;\;V)
\longrightarrow {\bf gr}^pC^{p+q}_{\rm BRST}
\end{equation}
\end{enumerate}
With these two observations, the Koszul isomorphisms:
\begin{align}
H^q({\cal L}^L;\; V) \simeq \;& H^q(Q^L_{\rm BRST};\;V)
\\    
H^p({\cal L}^R;\; {\bf C}) \simeq \;& H^p(Q^R_{\rm BRST};\;{\bf C}) 
=\mbox{Fun}(\lambda_R^{\otimes p})
\end{align}
imply that ${\bf gr}^pR:\;{\bf gr}^pC^{p+q}({\cal L}^{\rm tot}\;;\; {\bf g}_{\bar{0}}\;;\;V)\longrightarrow {\bf gr}^pC^{p+q}_{\rm BRST}$ is a 
quasi-isomorphism, {\it i.e.} $R$ is a filtered quasi-isomorphism. 

\subsection{An analogue of the Koszul resolution}
In fact, it is possible to glue two Koszul resolutions (one for ${\cal L}^L$ and 
another for ${\cal L}^R$) along ${\bf g}_{\bar{0}}$, as we will now explain\footnote{Note in the revised
version: we are greatful to the referee of \cite{Chandia:2013kja} 
for pointing out an error in the original version of this subsection}. 
Similarly to (\ref{KoszulSequence}), consider the following BRST-type complex: 
\begin{align}
0\longrightarrow {\bf C}
\longrightarrow
\mbox{Hom}_{{\bf g}_{\bar{0}}}(U{\cal L}^{\rm tot},{\bf C})
\longrightarrow
\mbox{Hom}_{{\bf g}_{\bar{0}}}(U{\cal L}^{\rm tot},{\cal P}^1) 
\longrightarrow \ldots
\label{RelativeInjectiveResolution} \\  
\ldots \longrightarrow 
\mbox{Hom}_{{\bf g}_{\bar{0}}}(U{\cal L}^{\rm tot},{\cal P}^n)
\longrightarrow
\mbox{Hom}_{{\bf g}_{\bar{0}}}(U{\cal L}^{\rm tot},{\cal P}^{n+1})
\longrightarrow \ldots
\nonumber
\end{align}
where the differential acts as follows:
\begin{align}\label{DifferentialInGluedComplex}
   d\phi = \;&
\mu_{\cal P}(\lambda_L^{\alpha})   
   \circ \phi \circ
\mu^{\rm \tiny right}_{U{\cal L}^{\rm tot}}(\nabla^L_{\alpha})
 + 
\mu_{\cal P}(\lambda_R^{\dot{\alpha}})   
   \circ \phi \circ
\mu^{\rm \tiny right}_{U{\cal L}^{\rm tot}}(\nabla^R_{\dot{\alpha}})
\end{align}
(notations as in (\ref{KoszulSequence})), and $\mbox{Hom}_{{\bf g}_{\bar{0}}}$ means linear maps invariant under the 
following action of ${\bf g}_{\bar{0}}$:
\begin{align}
   (\eta . \phi)(x) = \phi(x\eta) + \eta^{[mn]}t_{[mn]}^0\phi
\end{align}
We will call the two terms on the right hand side of (\ref{DifferentialInGluedComplex}) $d_L\phi$ and $d_R\phi$. We 
will introduce the abbreviated notation for the terms of (\ref{RelativeInjectiveResolution}):

\begin{equation}
0\longrightarrow {\bf C}\longrightarrow X^0 \longrightarrow X^1\longrightarrow
\ldots
\end{equation}
There is a bigrading: $X^n = \bigoplus_{p+q = n} X^{p,q}$ where $X^{p,q} = \mbox{Hom}_{{\bf g}_{\bar{0}}}(U{\cal L}^{\rm tot},{\cal P}^{p,q})$;
notice that $d_L:X^{p,q}\to X^{p+1,q}$ and $d_R:X^{p,q}\to X^{p,q+1}$.

We will now prove that (\ref{RelativeInjectiveResolution}) is a  $(U{\cal L}^{\rm tot},U{\bf g}_{\bar{0}})$-injective
$(U{\cal L}^{\rm tot},U{\bf g}_{\bar{0}})$-exact resolution of ${\bf C}$ in the sense of \cite{MR0080654}. 

\paragraph     {Proof}
Being $(U{\cal L}^{\rm tot},U{\bf g}_{\bar{0}})$-injective follows from Section 1 of \cite{MR0080654} (Lemma 1). Note 
that every term of (\ref{RelativeInjectiveResolution}) is a direct sum of finite-dimensional representations 
of ${\bf g}_{\bar{0}}$. This implies that the kernel and the image of every differential is a 
direct ${\bf g}_{\bar{0}}$-submodule as required in \cite{MR0080654}. It remains to prove the exactness. 
We will prove the equivalent statement, that the cohomology of the truncated 
complex:
\begin{equation}
   0\longrightarrow X^0\longrightarrow X^1\longrightarrow \ldots
\end{equation}
is only nonzero in the zeroth term: $H^0 = {\bf C}$. We will use the spectral 
sequence of the bicomplex $d = d_L + d_R$. Let us first calculate the 
cohomology of $d_L$. We will ``normal order'' the elements of $U{\cal L}^{\rm tot}$ by putting 
elements of $U{\cal L}^R$ to the left and elements of $U{\cal L}^L$ to the right. This gives 
an isomorphism of linear spaces:
\begin{equation}\label{GradedAsLinearSpaces}
\mbox{Hom}_{{\bf g}_{\bar{0}}} \left(U{\cal L}^{\rm tot}\;,\;{\cal P}^{n-p,\;p}\right) 
\;=\;
\mbox{Hom}_{\bf C}\left(U{\cal L}^L\;,\;{\cal P}_L^{n-p}\right)
\otimes
\mbox{Hom}_{\bf C}\left(U{\cal L}^R\;,\;{\cal P}_R^p\right)
\end{equation}
The differential $d_L$ only acts on the $\mbox{Hom}_{\bf C}\left(U{\cal L}^L\;,\;{\cal P}_L^{n-p}\right)$, while
$\mbox{Hom}_{\bf C}\left(U{\cal L}^R\;,\;{\cal P}_R^p\right)$ is ``inert''. The action of the differential on 
$\mbox{Hom}_{\bf C}\left(U{\cal L}^L\;,\;{\cal P}_L^{n-p}\right)$  is the same as in the Koszul complex of $U{\cal L}^L$. 
Therefore
the cohomology of $d_L$ is $\mbox{Hom}_{\bf C}\left(U{\cal L}^R\;,\;{\cal P}_R^p\right)$. The action of $d_R$ on the 
cohomology of $d_L$ is the same as the action of the differential in the  Koszul
complex of $U{\cal L}^R$. Therefore $H(d_R,H(d_L)) = {\bf C}$, corresponding to 
constant $\phi$. This completes the proof.

\paragraph     {Corollary}
This means that for any $U{\cal L}^{\rm tot}$-module $W$, the $\mbox{Ext}_{(U{\cal L}^{\rm tot},U{\bf g}_{\bar{0}})}(W,{\bf C})$ can be 
computed as the cohomology of the following complex:
\begin{align}
\ldots & \longrightarrow
   \mbox{Hom}_{U{\cal L}^{\rm tot}}\left(
      W\;,\;
      \mbox{Hom}_{{\bf g}_{\bar{0}}}(U{\cal L}^{\rm tot},{\cal P}^n)
   \right)
\longrightarrow
\nonumber \\   
& \longrightarrow
   \mbox{Hom}_{U{\cal L}^{\rm tot}}\left(
      W\;,\;
      \mbox{Hom}_{{\bf g}_{\bar{0}}}(U{\cal L}^{\rm tot},{\cal P}^{n+1})
   \right)
\longrightarrow\ldots
\label{ExtFromGluedComplex}
\end{align}
As in Section \ref{sec:KoszulAndDef}, there is an isomorphism of complexes (\ref{ExtFromGluedComplex}) and (\ref{BRSTComplexGeneralRepresentation}): 
\begin{align}
 \mbox{Hom}_{U{\cal L}^{\rm tot}}
 \left(
    W\;,\;
    \mbox{Hom}_{{\bf g}_{\bar{0}}}(U{\cal L}^{\rm tot},{\cal P}^n)
 \right) 
\simeq\;&
\mbox{Hom}_{{\bf g}_{\bar{0}}}(W, {\cal P}^n)
\\    
f \mapsto
\;& [w\mapsto f(w)({\bf 1})]
\end{align}
If $W$ is semisimple as a representation of ${\bf g}_{\bar{0}}$, then this shows that  
$\mbox{Ext}_{(U{\cal L}^{\rm tot},U{\bf g}_{\bar{0}})}(W,{\bf C})$ can be identified with the cohomology of (\ref{BRSTComplexGeneralRepresentation}) for
$V=W'$. 

\paragraph     {Variation}
Similarly, we can consider the following projective resolution:
\begin{align}
\ldots\longrightarrow 
({\cal P}^{n+1})'\otimes_{{\bf g}_{\bar{0}}} U{\cal L}^{\rm tot}
\longrightarrow 
({\cal P}^n)'\otimes_{{\bf g}_{\bar{0}}}U{\cal L}^{\rm tot}
\longrightarrow\ldots
\label{RelativeProjectiveResolution} \\  
\ldots\longrightarrow 
({\cal P}^1)'\otimes_{{\bf g}_{\bar{0}}}U{\cal L}^{\rm tot}
\longrightarrow 
{\bf C}\otimes_{{\bf g}_{\bar{0}}} U{\cal L}^{\rm tot}
\longrightarrow {\bf C}\longrightarrow 0
\nonumber
\end{align}
where the differential acts as follows:
\begin{align}\label{DifferentialInGluedComplex}
   \partial (s\otimes \xi) = \;&
   (s\circ\mu_{\cal P}(\lambda_L^{\alpha}))\otimes\xi\nabla^L_{\alpha} +
   (s\circ\mu_{\cal P}(\lambda_R^{\dot{\alpha}}))\otimes\xi\nabla^R_{\dot{\alpha}}
\end{align}
This means that $\mbox{Ext}_{(U{\cal L}_{\rm tot}, U{\bf g}_{\bar{0}})}({\bf C},V)$ can be computed as the cohomology of the
following complex:
\begin{align}
\ldots & \longrightarrow
   \mbox{Hom}_{U{\cal L}^{\rm tot}}\left(
      ({\cal P}^n)'\otimes_{{\bf g}_{\bar{0}}}U{\cal L}^{\rm tot}\;,\;V
   \right)
\longrightarrow
\nonumber \\   
& \longrightarrow
   \mbox{Hom}_{U{\cal L}^{\rm tot}}\left(
      ({\cal P}^{n+1})'\otimes_{{\bf g}_{\bar{0}}}U{\cal L}^{\rm tot}\;,\;V
   \right)
\longrightarrow\ldots
\label{ExtFromGluedComplexProjective}
\end{align}
As in Section \ref{sec:KoszulAndDef}, there is an isomorphism of complexes (\ref{ExtFromGluedComplexProjective}) and (\ref{BRSTComplexGeneralRepresentation}): 
\begin{align}
  \mbox{Hom}_{U{\cal L}^{\rm tot}}(({\cal P}^n)'\otimes_{{\bf g}_{\bar{0}}}U{\cal L}^{\rm tot}\;,\;V) \simeq \;&
{\cal P}^n\otimes_{{\bf g}_{\bar{0}}}V 
\label{IsomorphismToPV}
\\    
f \mapsto \;& [\lambda\mapsto f(\lambda\otimes {\bf 1})]
\label{ExplicitIsomorphism}
\end{align}
The expression $[\lambda\mapsto f(\lambda\otimes {\bf 1})]$ on the right hand side of (\ref{ExplicitIsomorphism}) denotes an 
element of ${\cal P}^n\otimes_{{\bf g}_{\bar{0}}}V$, understood as a ${\bf g}_{\bar{0}}$-invariant polynomial function of 
pure spinors of the order $n$, whose value on a pair of pure spinors 
$\lambda = (\lambda_L,\lambda_R)$ is defined as follows. Since $\lambda$ can be interpreted 
as an element of $({\cal P}^n)'$, we can consider $\lambda\otimes{\bf 1}$ an element of $({\cal P}^n)'\otimes_{{\bf g}_{\bar{0}}}U{\cal L}^{\rm tot}$;
then we can act on it by $f\in  \mbox{Hom}_{U{\cal L}^{\rm tot}}(({\cal P}^n)'\otimes_{{\bf g}_{\bar{0}}}U{\cal L}^{\rm tot}\;,\;V)$.

Eq. (\ref{IsomorphismToPV}) is another proof of (\ref{RelativeCohomologyCoincidesWithBRST}).

\subsection{Reduction to the cohomology of the ideal $I\subset {\cal L}^{\rm tot}$}
The following construction works for an arbitrary completely reducible 
representation $A$ of ${\bf g}_{\bar{0}}$. Given such an $A$, let us consider $H^n(Q_{\rm BRST}\;;\;V)$ 
in the special case:
\begin{equation}\label{DefV}
   V = \mbox{Hom}_{\bf C}\left( 
      U{\bf g}\otimes_{{\bf g}_{\bar 0}} A \;,\; {\bf C}
   \right)
\end{equation}
According to Section \ref{sec:LieAlgebraCohomology} $H^n(Q_{\rm BRST}\;;\;V)$ is equivalent to $H^n({\cal L}^{\rm tot}\;;\;{\bf g}_{\bar{0}}\;;\;V)$,
which in the case (\ref{DefV}) is the same as $\mbox{Ext}^n_{(U{\cal L}^{\rm tot},\;U{\bf g}_{\bar{0}})}(U{\bf g}\otimes_{{\bf g}_{\bar{0}}} A\;;\;{\bf C})$ \cite{MR0080654}. 
Consider the following complex of $U{\cal L}^{\rm tot}$-modules:
\begin{align}
\ldots \longrightarrow U{\cal L}^{\rm tot}\otimes_{{\bf g}_0} 
(\Lambda^2 I \otimes_{\bf C} A)
\longrightarrow U{\cal L}^{\rm tot}\otimes_{{\bf g}_0} (I \otimes_{\bf C} A)
\longrightarrow \;&  
\nonumber \\    
\longrightarrow U{\cal L}^{\rm tot}\otimes_{{\bf g}_{\bar{0}}} A
\longrightarrow U{\bf g}\otimes_{{\bf g}_{\bar{0}}} A \longrightarrow \;& 0
\label{RelativeResolution}
\end{align}
Here the action of ${\bf g}_{\bar{0}}$ on $\Lambda^p I \otimes_{\bf C} A$ is the sum of the adjoint action on $I$ and
the action on $A$. The complex (\ref{RelativeResolution}) is a $(U{\cal L}^{\rm tot},\;U{\bf g}_{\bar{0}})$-projective and  
$(U{\cal L}^{\rm tot},\;U{\bf g}_{\bar{0}})$-exact resolution of $U{\bf g}\otimes_{{\bf g}_{\bar{0}}}A$ as a $U{\cal L}^{\rm tot}$-module, in the sense
of \cite{MR0080654}; see Appendix \ref{sec:Exactness}. Therefore:
\begin{equation}\label{CohomologyWithA}
    H^n\left({\cal L}^{\rm tot}\;;\;{\bf g}_{\bar{0}}\;;\;
\mbox{Hom}_{\bf C}\left( 
      U{\bf g}\otimes_{{\bf g}_{\bar 0}} A \;,\; {\bf C}
   \right)
\right) = \mbox{Hom}_{{\bf g}_{\bar{0}}}\left(A\;, H^n(I) \right)
\end{equation}
\label{sec:ReductionToIdeal}
\paragraph     {Geometrical interpretation}
Consider the case when $A$ is a finite-dimensional representation. With $V$ 
defined by (\ref{DefV}) the BRST complex of (\ref{BRSTComplexGeneralRepresentation}) is:
\begin{equation}
   \mbox{Hom}_{{\bf g}_{\bar{0}}} \left(
      U{\bf g}\otimes_{{\bf g}_{\bar 0}} A \;,\; {\cal P}^{\bullet}
   \right)
\end{equation}
Geometrically, this is the space of $A'$-valued functions $f_a(g,\lambda_3,\lambda_1)$ where
the index $a$ enumerates a basis of $A'$, such that for $h\in G_{\bar{0}}$:
\begin{align}
   f_a(hg,\; h\lambda_3h^{-1},\; h\lambda_1h^{-1}) =\;& 
   f_a(g,\;\lambda_3,\lambda_1)
\label{CovarianceOfF}
\\    
f_a(gh,\;\lambda_3,\lambda_1) =\; & f_b(g,\;\lambda_3,\lambda_1)\rho^b_a(h)
\label{RotationsOfF}
\end{align}
More precisely, this is the space of {\em Taylor series} of sections of the 
pure spinor bundle over $AdS_5\times S^5$; the universal enveloping algebra is the 
space of {\em finite} linear combinations, {\it i.e.} we do not care about the
convergence of the Taylor series $f$. Equation (\ref{CovarianceOfF}) says that $f$ is a section 
of a bundle over the homogeneous space. On the other hand, Eq. (\ref{RotationsOfF}) requires 
that $f$ transform in a fixed representation $A'$ under the group $G_{\bar{0}}$ of 
{\em global rotations} around $g = {\bf 1}$. 

The space of Taylor series, as a representation of the global rotations $G_{\bar{0}}$,
is the direct sum of infinitely many finite-dimensional representations:
\begin{equation}
   \mbox{Hom}_{{\bf g}_{\bar{0}}} \left(
      U{\bf g} \;,\; {\cal P}^{\bullet}
   \right) = \bigoplus_{A} A\otimes
   \mbox{Hom}_{{\bf g}_{\bar{0}}} \left(
      U{\bf g}\otimes_{{\bf g}_{\bar 0}} A \;,\; {\cal P}^{\bullet}
   \right)
\end{equation}
Therefore (\ref{CohomologyWithA}) implies that:
\begin{equation}
   H^n\left(\;
      Q_{\rm BRST}\;,\;
      (U{\bf g})' \;
   \right) = H^n(I)
\end{equation}

\paragraph     {Action of the global symmetries}
Notice that $\bf g$ naturally acts on $H^m(I)$. This corresponds to the right action
of $\bf g$ on the BRST complex (\ref{StandardBRSTComplex}), {\it i.e.} to the global symmetries of the
$AdS_5\times S^5$ sigma-model.

\subsection{Ghost number 1: global symmetry currents}
The elements of $H^1(Q_{\rm BRST}\;;\;(U{\bf g})') = H^1(I)$ correspond to the global 
symmetry currents of the $\sigma$-model \cite{Berkovits:2004jw,Berkovits:2004xu,Bedoya:2010qz}. 
There are finitely many global symmetries. We have:
\begin{equation}
   H^1(I) = \left({I\over [I,I]}\right)'
\end{equation}
We will now show that $I\over [I,I]$ is a {\em finite-dimensional} representation of ${\bf g}$,
actually the adjoint representation of $\bf g$.
 
\paragraph     {Special notations for summation over repeating indices.} 
As already introduced in (\ref{BasisOfPSU}), the index $m$ enumerates the basis of the 
vector representation of ${\bf g}_{\bar{0}} = so(1,4)\oplus so(5)$, and runs from $0$ to $9$; more 
precisely, $m\in\{0,\ldots,4\}$ enumerates vectors of $so(1,4)$, and $m\in \{5,\ldots,9\}$
vectors of $so(5)$. For a vector $v^m$ we denote:
\begin{equation}
   v^{\overline{m}} = 
   \left\{
      \begin{array}{c}
         v^m \mbox{ \small\tt if } m\in\{0,\ldots,1\} \cr
         -v^m \mbox{ \small\tt if } m\in\{5,\ldots,9\} \end{array}\right.
\end{equation}
For two vectors $v^m$ and $w^m$ we denote:
\begin{align}
   v^mw^m = \;& v^0w^0 - \sum_{i=1}^9v^iw^i
\nonumber\\    
   v^mw^{\overline{m}} = \;& v^0w^0 - \sum_{i=1}^4v^iw^i + \sum_{i=5}^9 v^iw^i
\end{align}

\paragraph     {Proposition.} As a representation of $\bf g$, $I\over [I,I]$ is generated by 
the following objects\footnote{The coefficient $1\over 10$ depends on the 
choice of
normalization for $\nabla_{\alpha}$; in our conventions $f_{\alpha\beta}{}^m = \Gamma^m_{\alpha\beta}$, and the projection $\mbox{pr}(\nabla_m)$ of $\nabla_m$ to $\bf g$ satisfies: $(\mbox{ad}_{{\rm pr}(\nabla_m)})^2|_{{\bf g}_{\bar{3}}}=1$ --- no summation over $m$.} :
\begin{align}
T^2_m = \;& \nabla_m^L - \nabla_m^R
\label{T2}\\    
T^0_{[mn]} =\;& [\nabla^L_m,\nabla^L_n] - [\nabla^R_m,\nabla^R_n]
\label{T0}\\   
Z^L_{\alpha} =\;& \nabla^L_{\alpha} 
- {1\over 10} 
[\;\nabla^L_{\overline{m}}\;,\;[\nabla^L_m\;,\; \nabla^L_{\alpha}]\;]
\label{ZL}\\    
Z^R_{\dot{\alpha}} =\;& \nabla^R_{\dot{\alpha}} 
- {1\over 10} 
[\;\nabla^R_{\overline{m}}\;,\;[\nabla^R_m\;,\; \nabla^R_{\dot{\alpha}}]\;]
\label{ZR}
\end{align}
Notice that $[(\nabla^L_m - \nabla^R_m)\;,\;(\nabla^L_n - \nabla^R_n)] \in [I,I]$ implies that:
\begin{equation}
[\nabla^L_m,\nabla^L_n] + [\nabla^R_m,\nabla^R_n] - 2t_{[mn]}^{0} = 0\;\; \mbox{ mod } [I,I]
\end{equation}
Similarly, $[(\nabla_m^L - \nabla_m^R),[(\nabla^L_{\overline{m}} - \nabla^L_{\overline{m}}),
\nabla_{\alpha}^L]] \in [I,I]$ implies that:
\begin{align}
\nabla_{\alpha}^L - 
{1\over 10} f_{\alpha}{}^{m\dot{\alpha}}
[\nabla_{\overline{m}}^R,\nabla_{\dot{\alpha}}^R] = - Z^L_{\alpha} \mbox{ mod } [I,I]
\end{align}
We will write ``$\equiv 0$'' instead of ``$=0\mbox{ mod } [I,I]$''.

The $(30|32)$-dimensional linear space generated by $T_m^2,T_{[mn]}^0,Z_{\alpha}^L,Z_{\dot{\alpha}}^R$ is closed
under the action of ${\bf g}$. It must be the adjoint representation of $\bf g$. For 
example, let us consider $\{\nabla^L_{\alpha}\;,\;Z^L_{\beta}\}$. Modulo $[I,I]$ this is same as
$\{[\nabla^R_m\;,\;\nabla^R_{\dot{\alpha}}]\;,\;Z^L_{\beta}\}$, and using (\ref{CollapsToT0}), (\ref{ActionOfT0OnLeft}) and (\ref{ActionOfT0OnRight}) this is proportional to 
$T_m$. 

\paragraph     {Proof of the proposition.}
Let $J$ denote the subspace of $I/[I,I]$ generated by the action of $\bf g$ on (\ref{T2}), 
(\ref{T0}), (\ref{ZL}) and (\ref{ZR}). We have to prove that $J=I$. Let us consider some 
linear combination of commutators of $\nabla_{\alpha}^L$, for example:
\begin{align}\label{SumOfNestedCommutators}
\sum_{\vec{\alpha}}C^{\alpha_1\ldots\alpha_q}
   [\nabla^L_{\alpha_1},\{\nabla^L_{\alpha_2},\ldots
   [\nabla^L_{\alpha_{q-2}},\{\nabla^L_{\alpha_{q-1}},\nabla^L_{\alpha_q}\}]\ldots\}]
\end{align}
Suppose that the coefficients $C$ are such that this expression belongs to $I$.
We will prove that it also belongs to $J$, using the induction in $q$ --- the
number of commutators. Suppose that for $q<n$, all such expressions lie in 
$J$. We will prove that for $q=n$, (\ref{SumOfNestedCommutators}) is also in $J$.

Notice that:
\begin{equation}
\sum_{\vec{\alpha}}C^{\alpha_1\ldots\alpha_5}
   [\nabla^L_{\alpha_1},\{\nabla^L_{\alpha_2},\ldots
   \{\nabla^L_{\alpha_{q-1}},
   \left(\nabla^L_{\alpha_q} - 
      {1\over 10}f_{\alpha_q}{}^{\overline{m}\beta}[\nabla_m^R,\nabla^R_{\dot{\beta}}] 
   \right)
\}]\ldots\}] \in J
\end{equation}
because $\nabla^L_{\alpha} - {1\over 10}f_{\alpha}{}^{\overline{m}\beta}[\nabla_m^R,\nabla^R_{\dot{\beta}}]\in J$. Therefore, it remains to prove that the
following expression belongs to $J$:
\begin{equation}
\sum_{\vec{\alpha}}C^{\alpha_1\ldots\alpha_5}
   [\nabla^L_{\alpha_1},\{\nabla^L_{\alpha_2},\ldots
   \{\nabla^L_{\alpha_{q-1}},
   f_{\alpha_q}{}^{\overline{m}\beta}[\nabla_m^R,\nabla^R_{\dot{\beta}}] 
\}]\}] 
\end{equation}
(notice that it automatically belongs to $I$). When we commute $\nabla^R$ with $\nabla^L$, 
the number of commutators drops and we are left with $q-4$ commutators. This 
provides the step of the induction.

\paragraph     {Calculation of $\{\nabla_{\alpha}^L\;,\; Z^R_{\dot{\alpha}}\}$ and $\{\nabla_{\dot{\alpha}}^R\;,\;Z^L_{\alpha}\}$.}
Here we will prove that both $\{\nabla^L_{\alpha}\;,\; Z^R_{\dot{\alpha}}\}$ and $\{\nabla^R_{\dot{\alpha}}\;,\;Z^L_{\alpha}\}$ are proportional to
$f_{\alpha\dot{\alpha}}{}^{[mn]}T^0_{[mn]}$, and $[\nabla_m,T^2_n]$ is proportional to $f_{mn}{}^{[pq]}  T^0_{[pq]}$. 
Let us define $\nabla^R_{\alpha}$ and $\nabla^L_{\dot{\alpha}}$ so that:
\begin{align}
[\nabla_m^L,\nabla_{\alpha}^L] =\;& 
f_{m\alpha}{}^{\dot{\alpha}}\nabla^L_{\dot{\alpha}}
\label{DefNablaLDot}
\\       
[\nabla_m^R,\nabla_{\dot{\alpha}}^R] = \;&
f_{m\dot{\alpha}}{}^{\alpha}\nabla^R_{\alpha}
\label{DefNablaR}
\end{align}
That the RHS of (\ref{DefNablaLDot}) is proportional to $f_{m\alpha}{}^{\dot{\alpha}}$ and the RHS of (\ref{DefNablaR}) is 
proportional to $f_{m\dot{\alpha}}{}^{\alpha}$ follows from (\ref{DefNablaML}) and (\ref{DefNablaMR}). 

To calculate $\{\nabla^L_{\alpha}\;,\; Z^R_{\dot{\alpha}}\}$, $\{\nabla^R_{\dot{\alpha}}\;,\;Z^L_{\alpha}\}$ and $[\nabla_m,T^2_n]$ we start with the 
following observation:
\begin{equation}\label{NablaLZRPlusNablaRZL}
  \{\nabla_{\alpha}^L\;,\;Z^R_{\dot{\alpha}}\} 
+ \{\nabla_{\dot{\alpha}}^R\;,\;Z_{\alpha}^L\} \equiv 0
\end{equation}
This follows from:
\begin{equation}
0 \equiv \{\nabla^L_{\alpha} - \nabla^R_{\alpha}\;,\; 
\nabla^L_{\dot{\alpha}} - \nabla^R_{\dot{\alpha}}\} =
\{\nabla^L_{\alpha}\;,\;\nabla^L_{\dot{\alpha}}\} + 
\{\nabla^R_{\alpha}\;,\;\nabla^R_{\dot{\alpha}}\} - 2t^0_{\alpha\dot{\alpha}} 
\end{equation}
Also notice:
\begin{align}
&   \{[\nabla_m^L,\nabla_{\dot{\beta}}^L]\;,\; \nabla^R_{\dot{\alpha}} - \nabla^L_{\dot{\alpha}}\} \;\equiv
f_{m\dot{\beta}}{}^{\beta}\{\nabla^L_{\beta}\;,\; \nabla^R_{\dot{\alpha}} -\nabla^L_{\dot{\alpha}}\} \;=
\nonumber \\  
=\;& [\nabla_m^L\;,\;\{\nabla_{\dot{\beta}}^L\;,\; \nabla^R_{\dot{\alpha}} - \nabla^L_{\dot{\alpha}}\}] \;-\;
\{\nabla_{\dot{\beta}}^L\;,\; [\nabla_m^L\;,\;\nabla^R_{\dot{\alpha}} - \nabla^L_{\dot{\alpha}}]\} \; \equiv
\nonumber \\   
\equiv \;& 
- f_{\dot{\alpha}\dot{\beta}}{}^n[\nabla_m^L\;,\;\nabla_n^L-\nabla_n^R] \;-\;
f_{m\dot{\alpha}}{}^{\gamma}\{\nabla^L_{\dot{\beta}}\;,\; 
\nabla^R_{\gamma} - \nabla^L_{\gamma}\}
\end{align}
This implies:
\begin{align}
f_{m\dot{\beta}}{}^{\beta}\{\nabla^L_{\beta}\;,\;Z^R_{\dot{\alpha}}\} -
f_{m\dot{\alpha}}{}^{\gamma}\{\nabla_{\dot{\beta}}^R\;,\;Z^L_{\gamma}\} =
- f_{\dot{\alpha}\dot{\beta}}{}^n[\nabla_m^L\;,\;\nabla_n^L-\nabla_n^R] 
\end{align}
Similarly:
\begin{equation}
f_{m\beta}{}^{\dot{\beta}}\{\nabla_{\dot{\beta}}^R\;,\;Z_{\alpha}^L\}
- f_{m\alpha}{}^{\dot{\gamma}}\{\nabla^L_{\beta}\;,\;Z_{\dot{\gamma}}^R\} =
- f_{\alpha\beta}{}^n[\nabla_m^R\;,\;\nabla_n^R-\nabla_n^L] 
\end{equation}
Taking into account (\ref{NablaLZRPlusNablaRZL}), we get the following system of equations for 
$X_{\alpha\dot{\alpha}} = \{\nabla^L_{\alpha}\;,\;Z^R_{\dot{\alpha}}\}$ and $X_{mn} = [\nabla_m^L\;,\;\nabla_n^L - \nabla_n^R]$:
\begin{align}
2f_{m(\dot{\alpha}|}{}^{\gamma}X_{\gamma|\dot{\beta})} \;+\;
f_{\dot{\alpha}\dot{\beta}}{}^n X_{mn} \;=\;0
\label{Cocycle1}
\\    
2f_{m(\alpha}{}^{\dot{\gamma}}X_{\beta)\dot{\gamma}} \;+\;
f_{\alpha\beta}{}^n X_{mn}\;=\; 0
\label{Cocycle2}
\end{align}
This system of equations has the following solution, which defines $T^0_{[pq]}$:
\begin{align}
X_{\alpha\dot{\alpha}} =\;& f_{\alpha\dot{\alpha}}{}^{[pq]}\;T^0_{[pq]}
\label{DefT0}
\\    
X_{mn} = \;& - f_{mn}{}^{[pq]} \;T^0_{[pq]}
\end{align}
We have to prove that there are no other solutions. Let us use the
identity:
\begin{equation}
   f_m{}^{\alpha\beta}f_{\alpha\beta}{}^n = 16\; \delta_m^{\overline{n}}
\end{equation}
Contracting (\ref{Cocycle1}) and (\ref{Cocycle2}) with $f^{\dot{\alpha}\dot{\beta}}{}_{\overline{k}}$ and $f^{\alpha\beta}{}_{\overline{k}}$ we get:
\begin{align}
2f_{m\dot{\alpha}}{}^{\gamma}f_{\overline{k}}{}^{\dot{\alpha}\dot{\beta}}X_{\gamma\dot{\beta}} \;+\;
16\;  X_{mk} \;=\;& 0
\label{XgreekVsXlat1}
\\     
2f_{m\alpha}{}^{\dot{\gamma}}f_{\overline{k}}{}^{\alpha\beta} X_{\beta\dot{\gamma}} \;+\;
16\; X_{mk} \;=\;& 0
\end{align}
This implies:
\begin{equation}\label{ffX}
f_{m\dot{\alpha}}{}^{\gamma}f_{\overline{k}}{}^{\dot{\alpha}\dot{\beta}}X_{\gamma\dot{\beta}}
+
f_{\overline{k}\dot{\alpha}}{}^{\gamma}f_{m}{}^{\dot{\alpha}\dot{\beta}}X_{\gamma\dot{\beta}}
=0
\end{equation}
Let us assume that the pair $(m,k)$ is such that:\\
\begin{tabular}{rl}
either & $m\in \{0,\ldots,4\}$ and $k\in \{5,\ldots,9\}$ \\ 
or & $m\in \{5,\ldots,9\}$ and $k\in \{0,\ldots,4\}$;
\end{tabular}\\
then (\ref{ffX}) implies that for such pairs $(m,k)$ the expression $f_{m\dot{\alpha}}{}^{\gamma}f_{\overline{k}}{}^{\dot{\alpha}\dot{\beta}}X_{\gamma\dot{\beta}}$ is
symmetric under the exchange $m\leftrightarrow k$. But $X_{mk}$ is always antisymmetric under 
such an exchange. Therefore Eq. (\ref{XgreekVsXlat1}) implies that $X_{mk}$ is only nonzero when 
either both $m$ and $k$ belong to $\{0,\ldots,4\}$, or both $m$ and $k$ belong to 
$\{5,\ldots,9\}$. This means that $X_{mk}$ is proportional to $f_{mk}{}^{\bullet}$, and we can define
$T^0_{[pq]}$ from (\ref{DefT0}). Then (\ref{Cocycle1}) gives:
\begin{equation}
2f_{m(\dot{\alpha}|}{}^{\gamma}\left(
   X_{\gamma|\dot{\beta})} \;-\;f_{\gamma|\dot{\beta})}{}^{[pq]}Y_{[pq]}
\right) = 0
\end{equation}
which implies that $X_{\alpha\dot{\alpha}} \;=\;f_{\alpha\dot{\alpha}}{}^{[pq]}Y_{[pq]}$. 

\paragraph     {To summarize,} $I\over [I,I]$ is a finite-dimensional space, the adjoint 
representation of ${\bf g}$.

\subsection{Ghost number 2: vertex operators}
The cohomology group $H^n(I)$ is a linear space dual\footnote{This is the Poincar\'e duality, Section VI.3  of \cite{Knapp}.} to the homology $H_n(I)$.
The vertex operators correspond to $H^2(I) = (H_2(I))'$.
The linear space $H_2(I)$ consists of the expressions of the form:
\begin{align}
a = \;& \sum_i x_i\wedge y_i
\\    
\;& \sum_i [x_i,y_i]=0
\label{ZeroInternalCommutator}
\end{align}
where $x_i$ and $y_i$ are elements of $I$, with the equivalence relations:
\begin{equation}
   a\;\; \simeq\;\; a + [x,y]\wedge z + [y,z]\wedge x + [z,x]\wedge y
\end{equation}
We do not have the complete analysis at the ghost number two. It must be true 
that $H_2(I)$ correspond to the space of gauge-invariant\footnote{Gauge invariance means is the diffeomorphism invariance plus various gauge symmetries of the Type IIB SUGRA} operators at a marked 
point in $AdS_5\times S^5$. This is an infinite-dimensional representation of $\bf g$. 
The simplest element of $H_2(I)$ is:
\begin{align}\label{DilatonAtZero}
   {\cal O} = C^{\alpha\dot{\alpha}}(\nabla_{\alpha}^L - W^R_{\alpha})\wedge
(\nabla_{\dot{\alpha}}^R - W^L_{\dot{\alpha}})
\end{align}
This probably corresponds to the value of the dilaton\footnote{We did not 
prove that (\ref{DilatonAtZero}) is not exact. One can compute its value
on some vertex operator and show that it it nonzero; but this is technically
a nontrivial computation, and we did not do it}. It should be possible
to obtain other fields by acting on (\ref{DilatonAtZero}) with $\nabla_{\alpha}^L$ and $\nabla_{\dot{\alpha}}^R$.

\section{Flat space limit}\label{sec:FlatSpaceLimit}
In this section we will study the cohomology of the BRST operator in flat 
space. 

\vspace{7pt}

\noindent In flat space ${\cal L}^{\rm tot} = {\cal L}^L \oplus {\cal L}^R$. The limit of the BRST complex (\ref{StandardBRSTComplex})
is:
\begin{equation}\label{Qflat}
Q_{\rm SUGRA} =
   \lambda^{\alpha}_L \left(
      {\partial\over\partial \theta_L^{\alpha}} +
      \Gamma^m_{\alpha\beta}\theta_L^{\beta}{\partial\over\partial x^m}
   \right)
+
   \lambda^{\hat{\alpha}}_R \left(
      {\partial\over\partial \theta_R^{\hat{\alpha}}} +
      \Gamma^m_{\hat{\alpha}\hat{\beta}}
      \theta_R^{\hat{\beta}}{\partial\over\partial x^m}
   \right)
\end{equation}
acting on functions of $\theta_L,\theta_R,x,\lambda_L,\lambda_R$. 

\subsection{Ghost number 1.} 
The space $I\over [I,I]$ is generated by $\nabla^L_m - \nabla^R_m$, $[\nabla_m^L\;,\;\nabla_n^L]$, $W^{\alpha}_L$ and $W^{\dot{\alpha}}_R$. We 
observe:
\begin{align}
[\nabla_m^L\;,\;\nabla_n^L] = \;& - [\nabla_m^R\;,\;\nabla_n^R] \;\mbox{ mod }\; [I,I]
\end{align}
As a representation of $\bf susy$, this space should be the  dual to 
${\bf susy}+{\bf Lorentz}$. We observe:
\begin{align}
\{\nabla_{(\alpha}\;,\; \Gamma_{\beta)\gamma}^m W_L^{\gamma}\} \equiv \;&
{1\over 2}\Gamma^n_{\alpha\beta}[\nabla_n^L\;,\;\nabla_m^L]
\label{FlatSpaceNablaW}
\end{align}
As explained in \cite{Mafra:2009wq}, Eq. (\ref{FlatSpaceNablaW}) implies that $\nabla_{\alpha}W_L^{\gamma}$ is proportional to 
$(\Gamma_{mn})^{\gamma}_{\alpha}[\nabla_n^L\;,\;\nabla_m^L]$. 

\subsection{Ghost number 2.} 
We do not have the complete analysis at the
ghost number two. The RR should correspond to $W_L^{\alpha}\wedge W_R^{\dot{\alpha}}$. The NSNS 3-form 
field strength $H=dB$ should correspond to: 
\begin{equation}\label{HCorrespondsTo}
H_{klm} = (\nabla_{[k}^L-\nabla_{[k}^R)\wedge [\nabla^L_l,\nabla^L_{m]}]
\end{equation}
The following expression
\begin{equation}\label{Curvature}
R'_{klmn} = [\nabla_k^L,\nabla_l^L]\wedge [\nabla_m^R,\nabla_n^R]
+ [\nabla_m^L,\nabla_n^L]\wedge [\nabla_k^R,\nabla_l^R]
\end{equation}
should correspond to a linear combination of the curvature tensor $R_{klmn}$ and 
the second derivatives of the dilaton --- see Eq. (\ref{IdentificationOfRPrime}). It satisfies the 
relations:
\begin{align}
   R'_{klmn} = \;& R'_{mnkl} = - R'_{lkmn} 
\label{RiemannSymmetries}
   \\    
   R'_{[klmn]} =\;& 0
\label{RiemannJacobi} \\    
\nabla_{[j} R'_{kl]mn} = \;& 0
\label{RiemannSecondJacobi}
\end{align}
Notice that $R'_{k[lmn]}=0$ follows from (\ref{RiemannSymmetries}) and (\ref{RiemannJacobi}). Eq. (\ref{RiemannSymmetries}) follows 
immediately from (\ref{Curvature}). Here is the proof of (\ref{RiemannJacobi}):
\begin{align}
   R'_{klmn} = \;& [\nabla^L_k\;,\; \nabla^L_l]\wedge 
[\nabla^R_m -\nabla^L_m\;,\;\nabla^R_n - \nabla^L_n] 
+ ((kl)\leftrightarrow (mn))
\nonumber    
\\   
\Rightarrow 
   R'_{[klmn]} = \;&2\; [\nabla^L_{[k}\;,\; \nabla^L_l]\wedge 
[\nabla^R_m -\nabla^L_m\;,\;\nabla^R_{n]} - \nabla^L_{n]}] \; \equiv
\nonumber 
\\   
\equiv\;& 4\; [[\nabla^L_{[k}\;,\; \nabla^L_l]\;,\;\nabla^R_m -\nabla^L_m]\wedge 
(\nabla^R_{n]} - \nabla^L_{n]}) \;=\;0
\end{align}
--- here $[[\nabla^L_{[k},\nabla^L_l],\nabla^L_{m]}]=0$ because of the Jacobi identity.
To prove (\ref{RiemannSecondJacobi}) we observe that when calculating $\nabla_j\phi$ for any element $\phi$ of
$H_2(I)$, we can use either $\nabla^L_j\phi$ or $\nabla^R_j\phi$. Since both terms on the right hand 
side of (\ref{Curvature}) are in $H_2(I)$, we are free to use $\nabla^L_j$ when calculating
$\nabla_j([\nabla_k^L,\nabla_l^L]\wedge [\nabla_m^R,\nabla_n^R])$ and $\nabla_j^R$ when calculating $\nabla_j([\nabla_m^L,\nabla_n^L]\wedge [\nabla_k^R,\nabla_l^R])$.
Those are both zero because of the Jacobi identity.

\paragraph     {Mismatch.} It turns out that the linearized SUGRA equations 
of motion are not satisfied, because  $\nabla^k H_{klm}\neq 0$. Using the identities from
Appendix B of \cite{Mafra:2009wq}, we derive using (\ref{HCorrespondsTo}):
\begin{align}
   \nabla^kH_{klm} = \;& 
-{2\over 3}[\nabla_k^L,\nabla_l^L]\wedge [\nabla_k^L,\nabla_m^L] 
+ {1\over 3}(\nabla_k^L-\nabla_k^R)\wedge 
[\nabla_k^L,[\nabla_l^L,\nabla_m^L]]\;+
\nonumber \\    
& + {1\over 3}(\nabla_{[l}^L - \nabla_{[l}^R)\wedge 
\Gamma_{m]\alpha\beta}\{W^{\alpha}_L,W^{\beta}_L\}
\end{align}
However, the derivatives of  $\nabla^k H_{klm}$ are all zero\footnote{Since the homology of $I$ is $I$-invariant, we can calculate either $\nabla_n^L\nabla^kH_{klm}$ or $\nabla_n^R\nabla^kH_{klm}$; it is easier to calculate  $\nabla_n^R\nabla^kH_{klm}$}: 
\begin{equation}\label{DerivativesOfDivHAreZero}
\nabla_n\nabla^kH_{klm} =0
\end{equation}
therefore this is a ``zero mode effect''. Moreover, we have:
\ifodd\amshow 
\fi
\begin{align}
   \nabla^kH_{klm} =\;& \nabla_{[l}A_{m]}^L = \nabla_{[l}A_{m]}^R
\label{DivHVsDA}\\    
\mbox{\tt\small where }\;& 
A^L_m = {2\over 3}(\nabla_n^L-\nabla_n^R)
\wedge [\nabla_n^L,\nabla_m^L] 
+ {1\over 3}\Gamma_{\alpha\beta m}W_L^{\alpha}\wedge W_L^{\beta}
\nonumber\\    
\;& A^R_m = {2\over 3}(\nabla_n^R-\nabla_n^L)
\wedge [\nabla_n^R,\nabla_m^R] 
+ {1\over 3}\Gamma_{\alpha\beta m}W_R^{\alpha}\wedge W_R^{\beta}
\end{align}
Notice that $A^L_m$ and $A^R_m$ are both in $H_2(I)$. 

\paragraph     {The dilaton}
The difference $A_m^L - A_m^R$ should be identified with the first derivative of 
the dilaton $\partial_m\phi$. Notice that:
\begin{equation}
   \nabla_n(A_m^L - A_m^R) = 
 {4\over 3} [\nabla_k^L,\nabla_{(m}^L]\wedge [\nabla_{n)}^R,\nabla_k^R]
\end{equation}
This is in agreement with the statement that (\ref{Curvature}) is a linear combination of 
the Riemann-Christoffel tensor $R_{klmn}$ and the derivatives of the dilaton 
$\partial_{[l}g_{k][m}\partial_{n]} \phi$. Indeed, we have:
\begin{align}
   g^{lm}\left(
[\nabla_k^L,\nabla_l^L]\wedge [\nabla_m^R,\nabla_n^R]
+ [\nabla_m^L,\nabla_n^L]\wedge [\nabla_k^R,\nabla_l^R]
\right) - {3\over 4} \nabla_n (A_k^L - A_k^R) = 0
\end{align}
which is the Einstein's equation $R_{kn} = 0$ for the Ricci tensor
$R_{kn}=g^{lm}R_{klmn}$, if we identify:
\begin{align}\label{IdentificationOfRPrime}
&   [\nabla_k^L,\nabla_l^L]\wedge [\nabla_m^R,\nabla_n^R]
+ [\nabla_m^L,\nabla_n^L]\wedge [\nabla_k^R,\nabla_l^R] =
\nonumber \\   
= \;& R_{klmn} + 
\partial_{[l}g_{k][m}\partial_{n]}\phi
\end{align}
\ifodd\amshow
\includegraphics[scale=0.5]{photos/ddPhi.png}\\ 
\fi
where $R_{klmn}$ is the Riemann-Christoffel tensor in the Einsten frame, and
$\partial_n\phi = {3\over 8} (A^L_n - A^R_n)$. Also observe that $\nabla_n(A_n^L - A_n^R) =0$ --- the 
Klein-Gordon equation for the dilaton. Indeed:
\begin{align}
&   [\nabla_k^L,\nabla_l^L]\wedge [\nabla_k^R,\nabla_l^R] =
\nonumber \\  
=\;& 
[(\nabla_k^L-\nabla_k^R),(\nabla_l^L-\nabla_l^R)]\wedge [\nabla_k^R,\nabla_l^R] 
\simeq
\nonumber \\   
\simeq \;& 
 2(\nabla_k^L - \nabla_k^R)\wedge [\nabla_l^R,[\nabla_l^R,\nabla_k^R]] =
- (\nabla_k^L - \nabla_k^R)\wedge \Gamma_{k\alpha\beta} \{W^{\alpha}_R, W^{\beta}_R\}
\simeq
\nonumber \\  
\simeq \;&
 \Gamma_{k\alpha\beta} [\nabla_k^R , W^{\alpha}_R ]\;\wedge
W^{\beta}_R = 0
\end{align}
(We used the Dirac equation $ \Gamma_{k\alpha\beta} [\nabla_k^R , W^{\alpha}_R ] = 0$.) 

\paragraph     {Unphysical operator}
We have seen that the difference $A_m^L - A_m^R$ corresponds to the derivative of 
the dilaton: $\partial_m\phi$. But the sum $A_m^L + A_m^R$ presents a problem. Observe that:
\begin{align}
   \nabla_l(A_m^L + A_m^R) =\;& \nabla_{[l}(A_{m]}^L + A_{m]}^R)
\\     
   \nabla_k\nabla_l(A_m^L + A_m^R) =\;& 0
\end{align}
This means that the first derivative of $(A_m^L + A_m^R)$ is {\em a constant}.

\paragraph     {Relation to the results of \cite{Bedoya:2010qz,Mikhailov:2012id}} 
This mismatch is not surprizing. We know from \cite{Bedoya:2010qz} that the zero momentum states
are not correctly reproduced as the cohomology of the ``naive'' BRST complex 
(\ref{StandardBRSTComplex}). Therefore we do expect a mismatch in the zero mode sector of the space 
of local operators. 

A state on which $A_m^L + A_m^R$ is nonzero is described in \cite{Mikhailov:2012id}. It is obtained as 
the flat space limit of the nonphysical AdS vertex of \cite{Bedoya:2010qz} with the internal 
commutator taking values in ${\bf g}_{\bar{2}}$ (using the notations of Section \ref{sec:BRSTComplexTypeIIB}). In this 
case $A_m^L + A_m^R$ is constant --- the gradient of the ``asymmetric dilaton''. 

Besides being constant,  $A_m^L + A_m^R$ can also be depending on $x$ linearly.
To obtain the state with  $A_m^L + A_m^R$ depending linearly on $x$, we have to 
consider the flat space limit of the nonphysical vertex ${\cal B}_{ab}j^a\wedge j^b$ with the 
internal commutator $f^{ab}{}_c {\cal B}_{ab}$ taking values in ${\bf g}_{\bar{0}}$ \cite{Bedoya:2010qz,Mikhailov:2012id}. It depends on a 
constant antisymmetric tensor $B_{mn}$. The leading term in the flat space limit 
is a trivial constant NSNS $B$-field $B_{mn}dx^m\wedge dx^n$, which can be gauged away.
Discarding the terms with $\theta$'s, the leading nontrivial term is:
\begin{equation}
   B_{mn}dx^m\wedge 
\left(x^n\sum_{k=0}^4(dx_kx^k) - dx^n\sum_{k=0}^4(x_k x^k)\right)
\end{equation}
This does not solve the SUGRA equations $\partial^nH_{nml}=0$, instead $\partial^nH_{nml}$ is 
proportional to $B_{mn}dx^m\wedge dx^n$ --- a constant 2-form. 

\ifodd\amshow
\fi

In terms of the unintegrated vertex, the observable $A_m^L + A_m^R$ should be 
identified as follows. It is proportional to $\partial^nB_{mn}$ in the gauge where the 
vertex has ghost number $(1,1)$, {\it i.e.}  only $\lambda_L\lambda_R$ terms, no $\lambda_L\lambda_L$ and 
$\lambda_R\lambda_R$ terms\footnote{If we try to change the gauge
$B_{mn}\rightarrow B_{mn} + \partial_{[m}\Lambda_{n]}$ to get rid of $\partial^nB_{mn}$, this would generate some 
$\lambda_L\lambda_L$ and $\lambda_R\lambda_R$ terms \cite{Bedoya:2010qz}.}. 

\paragraph     {Nonphysical operator: summary}

Let us denote:
\begin{align}
[\nabla_k^L,\nabla_l^L]\wedge [\nabla_m^R,\nabla_n^R]
+ [\nabla_m^L,\nabla_n^L]\wedge [\nabla_k^R,\nabla_l^R]
\;& = {\cal R}_{klmn}
\\   
(\nabla_{[k}^L-\nabla_{[k}^R)\wedge [\nabla^L_l,\nabla^L_{m]}] 
\;& = H_{klm} =\partial_{[k}B_{lm]}
\\  
A_m^{\pm}\;& = A_m^L \pm A_m^R
\label{DefApm}
\end{align}
We get the following equations of motion:
\begin{align}
   g^{lm}{\cal R}_{klmn} \;& = {3\over 4} \nabla_{(k}A^-_{n)}
\\   
0 \;& = \nabla_{[k}A^-_{n]}
\\   
\nabla^k H_{klm} \;& = \nabla_{[l} A^+_{m]}
\\  
0 \;& = \nabla_{(l}A^+_{m)}
\end{align}
The gradient of the dilaton corresponds to $A^-_n$, while $A_n^+$ does not have a 
clear interpretation in the Type IIB supergravity. The ``observable'' $A_n^+$ is
dual to the unphysical vertex of \cite{Mikhailov:2012id}. The unphysical vertex is not BRST 
trivial. However, as we explained in \cite{Mikhailov:2012id}, it should be thrown away because it 
leads to a quantum anomaly in the worldsheet sigma-model at the 1-loop level.

\paragraph     {Generic element of $H_2(I)$}
The ``generic'' element is:
\begin{equation}
   {\cal O} = x_L\wedge x_R
\end{equation}
where $x_L\in I\cap {\cal L}^L$ and $x_R\in I\cap {\cal L}^R$. Notice that the following expression:
\begin{equation}
   (\nabla_m x_L)\wedge x_R - x_L\wedge (\nabla_m x_R)
\end{equation}
is zero in homology, {\it i.e.} exact:
\begin{align}
     (\nabla_m x_L)\wedge x_R - x_L\wedge (\nabla_m x_R) =
\delta( (\nabla_m^L - \nabla_m^R)\wedge x_L \wedge x_R )
\end{align}
Indeed, the generic gauge-invariant SUGRA operator can be understood as the 
product of two gauge-invariant Maxwell operators ${\cal O}_L$ and ${\cal O}_R$, with the 
condition that ${\cal O}_L \;\stackrel{\leftrightarrow}{\partial\over{\partial x^m}} {\cal O}_R  = 0$. The zero momentum special operators of the
form (\ref{DivHVsDA}) are not of this form.

\subsection{Higher ghost numbers}
This section was {\bf added in the revised version} of the paper. 
We have previously claimed that the cohomology at the ghost number higher than 
2 vanishes. We are greateful to the referee for insisting that we present a
proof of this statement. Upon careful examination, it turns out that the 
statement is wrong. There is some nontrivial cohomology at least at the ghost 
number 3. Here we will only do a preliminary analysis:
\begin{itemize}
\item We prove that the cohomology at the ghost number $>4$ vanishes.
\item We give an example of the nontrivial cohomology class at the ghost number 3.
\end{itemize}
We suspect that the cohomology at the ghost numbers 3 and 4 is a
finite-dimensional space, and is in some way related to the unphysical states
of \cite{Bedoya:2010qz,Mikhailov:2012id}.

We will start by proving the vanishing theorem for the super-Maxwell cohomology
at the ghost number higher than 1. We will then point out that the SUGRA BRST 
complex is {\em amlost} the tensor product of two super-Maxwell complexes (the 
``left sector'' and the ``right sector''). 
If it were, literally, the tensor product, that would indeed imply the 
vanishing theorem at the ghost number $>2$. But in fact, even in flat space 
there is some ``interaction'' between the left and the right sector, and this 
leads to a nontrivial cohomology at least at the ghost number 3.

\subsubsection{Super-Maxwell BRST complex}
The cohomology of the super-Maxwell BRST complex:
\begin{equation}
   Q_{\rm SMaxw} = \lambda^{\alpha}\left(
      {\partial\over\partial \theta^{\alpha}} +
      \Gamma^m_{\alpha\beta}\theta^{\beta}{\partial\over\partial x^m}
   \right)
\end{equation}
is only nontrivial at the ghost numbers 0 and 1. 

\paragraph     {Sketch of the proof} This fact is well-known in the pure spinor 
formalism. At the ghost number 0, the cohomology is formed by the constants 
(no dependence on $\lambda$, $x$ and $\theta$). At the
ghost number 1, the cohomology is the solutions of the free Maxwell equation 
and the free Dirac equation. The vanishing of the cohomology at the ghost 
number 2 is equivalent to the following two statements: 1) for any current $j_m$ 
such that $\partial_mj_m=0$ always exists the gauge field $F_{mn}$ satisfying 
$\partial_{[k}F_{lm]}=0$ and $\partial_mF_{mn} = j_n$ and 2) for any spinor $\psi$ exists a spinor $\phi$ such 
that $\Gamma^m\partial_m\phi = \psi$. The vanishing of the cohomology at the ghost number 3 is 
equivalent to the statement that for any $\rho$ exists $j_m$ such that $\partial_m j_m = \rho$.
All these facts are proven in any graduate course of classical electrodynamics.

\subsubsection{Type IIB BRST complex}
The BRST complex of Type IIB in flat space is {\em almost} the tensor product
of two SMaxwell complexes:
\begin{equation}\label{BRSTSMaxwTimesSMaxw}
Q_{{\rm SMaxw}\otimes {\rm SMaxw}} =
   \lambda^{\alpha}_L \left(
      {\partial\over\partial \theta_L^{\alpha}} +
      \Gamma^m_{\alpha\beta}\theta_L^{\beta}{\partial\over\partial x_L^m}
   \right)
+
   \lambda^{\hat{\alpha}}_R \left(
      {\partial\over\partial \theta_R^{\hat{\alpha}}} +
      \Gamma^m_{\hat{\alpha}\hat{\beta}}
      \theta_R^{\hat{\beta}}{\partial\over\partial x_R^m}
   \right)
\end{equation}
The cohomology of (\ref{BRSTSMaxwTimesSMaxw}) is the tensor product of the cohomologies of two 
super-Maxwell complexes. Therefore it is only nontrivial at the ghost numbers 
0,1 and 2. However, in the Type IIB BRST complex there is no separation of $x$
into $x_L$ and $x_R$. The actual BRST complex is therefore different from (\ref{BRSTSMaxwTimesSMaxw}):
\begin{equation}\label{QSUGRA}
Q_{\rm SUGRA} =
   \lambda^{\alpha}_L \left(
      {\partial\over\partial \theta_L^{\alpha}} +
      \Gamma^m_{\alpha\beta}\theta_L^{\beta}{\partial\over\partial x^m}
   \right)
+
   \lambda^{\hat{\alpha}}_R \left(
      {\partial\over\partial \theta_R^{\hat{\alpha}}} +
      \Gamma^m_{\hat{\alpha}\hat{\beta}}
      \theta_R^{\hat{\beta}}{\partial\over\partial x^m}
   \right)
\end{equation}
The difference is that the left and the right sector have a common $x$ instead 
of separate $x_L$ and $x_R$. We also write:
\begin{equation}
   Q_{\rm SUGRA} = Q_L + Q_R
\end{equation}
where $Q_L$ and $Q_R$ are the first and second terms on the right hand side of
(\ref{QSUGRA}).

\paragraph     {Vanishing theorem:} $H^n_{Q_{\rm SUGRA}} = 0$ for $n>4$. 
Let us consider, for example, a vertex of the ghost number $5$. 

\paragraph     {Lemma} Given a vertex at the ghost number 5, we can always 
modify it by adding $Q$-exact terms so that the new vertex has only terms of 
the type $\lambda_L^1\lambda_R^4$. 

We have to prove that the terms with $\lambda_R^5$, $\lambda_L^2\lambda_R^3$, $\lambda_L^3\lambda_R^2$, $\lambda_L^4\lambda_R$ and $\lambda_L^5$ can be 
gauged away. The term with $\lambda_R^5$ is $Q_R$-closed. Suppose that the term with the
lowest power of $\theta_R$ is proportional to $\lambda_R^5\theta_R^p$. We observe that this term is
closed under $\lambda_R{\partial\over\partial\theta_R}$ and therefore is equal to $\lambda_R{\partial\over\partial\theta_R}$ of some expression 
proportional to $\lambda_R^4\theta_R^{p+1}$. This means that we can add $Q$-exact terms so that the
new vertex has terms of the order $\lambda_R^5$ starting with $\lambda_R^5\theta_R^{p+2}$. An induction by 
$p$ implies that the terms containing $\lambda_R^5$ can be all gauged away. Similarly, we 
can gauge away terms proportional to $\lambda_L^5$, then terms proportional to $\lambda_L^4\lambda_R$, 
then $\lambda_L^3\lambda_R^2$, then $\lambda_L^2\lambda_R^3$. This {\bf proves the Lemma}.

Now we are left with the terms proportional to $\lambda_L^1\lambda_R^4$. In this gauge the vertex
operator is both $Q_R$-closed and $Q_L$-closed. Let us look at the expansion in
powers of $\theta_R$. Schematically:
\begin{equation}
   V = \lambda_R^4\left(
      \theta_R^k\phi_k(\lambda_L,\theta_L,x) + 
      \theta_R^{k+1}\phi_{k+1}(\lambda_L,\theta_L,x) + \ldots
\right)
\end{equation}
were every $\phi_j$ is linear in $\lambda_L$. We observe that all these $\phi_j$s are annihilated
by $Q_L$ (because $Q_LV=0$ and $Q_L$ does not act on $\theta_R$):
\begin{equation}
   Q_L\phi_j =0
\end{equation}
We also observe that in the leading term, the coefficient of $\phi_k$ is annihilated
by $\lambda_R{\partial\over\partial\theta_R}$. This implies:
\begin{align}
   V =\;& Q_{\rm SUGRA}\left(
      \lambda_R^3\theta_R^{k+1}\phi_k(\lambda_L,\theta_L,x)
   \right) \; + 
\nonumber \\
\;& + \lambda_R^4\left(
      \theta_R^{k+1}\phi_{k+1}(\lambda_L,\theta_L,x) + 
      \theta_R^{k+2}\tilde{\phi}_{k+2}(\lambda_L,\theta_L,x) + \ldots
   \right)
\end{align}
This means that we are able to increase the order of the leading term by
adding a $Q_{\rm SUGRA}$-exact expression. The induction in $k$ 
{\bf proves the Theorem}.

But is it true that $H^n_{Q_{\rm SUGRA}} =0 $ for $n=3$ and $n=4$? It turns out that 
at least for $n=3$ the cohomology is nontrivial.  
The fact that the cohomology at the ghost number higher than 2 is
nontrivial is (for us) unexpected. We will leave this for future research,
giving here only an example.

\paragraph     {Example of a vertex at the ghost number 3}
For any constant 5-form $F$, let us denote $\hat{F} = F_{klmnp}\Gamma^{klmnp}$. Consider the 
following coboundary of $Q_{{\rm SMaxw}\otimes {\rm SMaxw}}$:
\begin{align}\label{PhiIsQOfPsi}
   \Phi[F] = \;& Q_{{\rm SMaxw}\otimes {\rm SMaxw}}\Psi[F]
\end{align}
where
\begin{align}
\Psi[F] =\;&
      (\theta_L\Gamma^p\lambda_L) \left(\theta_L\Gamma_p\;
   (x_L^m\Gamma_mx_R^n\Gamma_n + 5||x_L||^2)\hat{F}\;
      \Gamma_q\theta_R
   \right)(\lambda_R\Gamma^q\theta_R)\;+
\nonumber \\  
\;& +       (\theta_L\Gamma^p\lambda_L) \left(\theta_L\Gamma_p\;
x_L^m\Gamma_m f[\lambda_R\theta_R^4]\right) 
+ \left(
   g_n[\lambda_L\theta_L^4]x_R^n\hat{F}\Gamma_q\theta_R
\right)(\lambda_R\Gamma^q\theta_R)
\end{align}
where $f[\lambda_R\theta_R^4]$ is chosen so that:
\begin{equation}
   \left(\lambda_R{\partial\over\partial\theta_R} + (\theta_R\Gamma^l\lambda_R)
{\partial\over\partial x_R^l}\right)
\left(
   x_R^n\Gamma_n \hat{F} \Gamma_q\theta_R(\lambda_R\Gamma^q\theta_R)
   + f[\lambda_R\theta_R^4]
\right)= 0
\end{equation}
and $g[\lambda_L\theta_L^4]$ is chosen so that:
\begin{align}
\left(\lambda_L{\partial\over\partial\theta_L} + (\theta_L\Gamma^l\lambda_L)
{\partial\over\partial x_L^l}\right)\left(
(\theta_L\Gamma^p\lambda_L)
   \theta_L\Gamma_p \left(x_L^m\Gamma_m \Gamma_n - 10 x_L^n\right)
   + g^n[\lambda_L\theta_L^4]
\right) = 0
\end{align}
Such $f[\lambda_R\theta_R^4]$ and $g^n[\lambda_L\theta_L^4]$ exist because the expression $x_R^n\Gamma_n \hat{F}$ satisfies the
``right'' Dirac equation:
\begin{equation}
{\partial\over\partial x_R^k}\left(x_R^n\Gamma_n \hat{F}\right)\Gamma_k = 0
\end{equation}
and the expression $\left(x_L^m\Gamma_m \Gamma_n - 10 x_L^n\right)$ satisfies the ``left'' Dirac equation:
\begin{equation}
   {\partial\over\partial x^k_L}\Gamma_k \left(x_L^m\Gamma_m \Gamma_n - 10 x_L^n\right) = 0
\end{equation} 
We will now prove that $\Phi[F]$ depends on $x_L$ and $x_R$ only in the combination
$x_L + x_R$. Indeed, for a constant $c^m$ let us introduce $\Xi[c,F]$ as follows:
\begin{align}
\Xi[c,F]=\;&   
c^m\left(
   {\partial\over\partial x_L^m} - {\partial\over\partial x_R^m}
\right) \Psi[F] =
\nonumber \\  
=\;& 
(\theta_L\Gamma^p\lambda_L)\left(
   \theta_L\Gamma_p \;c^m\Gamma_m \; x_R^n\Gamma_n \hat{F}\Gamma_q\theta_R
\right)(\lambda_R\Gamma^q\theta_R) \;+
\nonumber \\   
\;& 
+ (\theta_L\Gamma^p\lambda_L) \left(\theta_L\Gamma_p\;
c^m\Gamma_m f[\lambda_R\theta_R^4]\right) \; -
\nonumber \\    
\;& 
- (\theta_L\Gamma^p\lambda_L) \left(\theta_L\Gamma_p\;
   (x_L^m\Gamma_mc^n\Gamma_n - 10(x_Lc))\hat{F}\;
      \Gamma_q\theta_R
   \right)(\lambda_R\Gamma^q\theta_R) \; -
\nonumber \\  
\;&
- \left(g^n[\lambda_L\theta^4_L] c_n\hat{F}\;
      \Gamma_q\theta_R
   \right)(\lambda_R\Gamma^p\theta_R) 
\label{XicF}
\end{align}
and we observe that:
\begin{equation}\label{QdIsZero}
   Q_{{\rm SMaxw}\otimes {\rm SMaxw}}\;
c^m\left(
   {\partial\over\partial x_L^m} - {\partial\over\partial x_R^m}
\right) \Psi[F] = 0
\end{equation}
Since $Q_{{\rm SMaxw}\otimes {\rm SMaxw}}$ commutes with $c^m\left(
   {\partial\over\partial x_L^m} - {\partial\over\partial x_R^m}
\right)$, Eq. (\ref{QdIsZero}) implies that
$\Phi[F]$ depends on $x_L$ and $x_R$ only in the combination $x_L + x_R$, and is therefore
a cocycle of $Q_{\rm SUGRA}$. We will now prove that $\Phi[F]$ is not a coboundary of
$Q_{\rm SUGRA}$. We know that $\Phi[F]$ {\em is} a coboundary of $Q_{{\rm SMaxw}\otimes {\rm SMaxw}}$, 
{\it i.e.} once we introduce separate $x_L$ and $x_R$ we have (\ref{PhiIsQOfPsi}). 
The question is: 

\parbox[t][5em][c]{0.80\textwidth}{can we modify $\Psi[F]$, by adding to it something 
$Q_{{\rm SMaxw}\otimes {\rm SMaxw}}$-closed, so that the modified $\Psi[F]$ is annihilated by 
${\partial\over\partial x_L} - {\partial\over\partial x_R}$?
}
\parbox[t][3em][t]{0.15\textwidth}{
\begin{equation}\label{Question}
\;
\end{equation}}

\noindent
In order to answer this question, it is useful to consider $c$ as a ghost and
interpret $\Xi[c,F]$ as a cocycle of the nilpotent operator $c^m\left(
   {\partial\over\partial x_L^m} - {\partial\over\partial x_R^m}
\right)$ 
acting {\em on the cohomology} of $Q_{{\rm SMaxw}\otimes {\rm SMaxw}}$. The answer to the question 
(\ref{Question}) is positive only if $\Xi[c,F]$ is a coboundary in this complex. The 
cohomology of $Q_{{\rm SMaxw}\otimes {\rm SMaxw}}$ is the tensor product of two super-Maxwell 
solutions. We will now prove that $\Xi[c,F]$ represents a nonzero element of:
\begin{equation}\label{GroupH1}
   H^1\left(\; c^m\left(
         {\partial\over\partial x_L^m} - {\partial\over\partial x_R^m}
      \right)\;,\;\; {\rm SMaxw}_{(x_L)}\otimes {\rm SMaxw}_{(x_R)}
\right)
\end{equation}
Remember that super-Maxwell is a direct sum of a solution of the free Maxwell 
equations and a solution of the free Dirac equation. Looking at (\ref{XicF}), the 
corresponding cocycle corresponds to the tensor product of two solutions of 
the free Dirac equation. Such an element of $ {\rm SMaxw}_{(x_L)}\otimes {\rm SMaxw}_{(x_R)}$ can be 
represented as a bispinor field $\psi^{\alpha\hat{\beta}}(x_L,x_R)$ satisfying:
\begin{align}
   \Gamma^m_{\alpha\alpha'}{\partial\over\partial x^m_L}
   \psi^{\alpha'\dot{\beta}}(x_L,x_R) = \;& 0
\label{LeftDiracEqn}\\    
{\partial\over\partial x_R^m}\psi^{\alpha\dot{\beta}'}(x_L,x_R)
\Gamma^m_{\dot{\beta}'\dot{\beta}} = \;& 0
\label{RightDiracEqn}
\end{align}
The element of (\ref{GroupH1}) corresponding to $\Xi[c,F]$ is:
\begin{align}\label{Cocycle}
   \psi(c;x_L,x_R)^{\alpha\dot{\beta}} = \;& 
   \left(
      \hat{c} \hat{x}_R \hat{F} - (\hat{x}_L\hat{c} - 10 (x_L\cdot c))\hat{F}
   \right)^{\alpha\dot{\beta}}
\end{align}
where hat over letter stands for the contraction with the gamma-matrices,
{\it e.g.} $\hat{x}_R = \Gamma_m x^m_R$. Let us analize the possibility of (\ref{Cocycle}) being in the
image of $c^m\left(
   {\partial\over\partial x_L^m} - {\partial\over\partial x_R^m}
\right)$:
\begin{align}
\;&      \left(
      \hat{c} \hat{x}_R \hat{F} - (\hat{x}_L\hat{c} - 10 (x_L\cdot c))\hat{F}
   \right)^{\alpha\dot{\beta}}\; \stackrel{?}{=}
\nonumber \\    
\stackrel{?}{=}\;&
c^m\left(
   {\partial\over\partial x_L^m} - {\partial\over\partial x_R^m}
\right)
\left(
   \phi^{\alpha\dot{\beta}}_{mn}x_L^mx_L^n  + 
   \chi^{\alpha\dot{\beta}}_{mn}x_L^mx_R^n  + 
   \sigma^{\alpha\dot{\beta}}_{mn}x_R^mx_R^n
\right)
\end{align}
with all three $\phi^{\alpha\dot{\beta}}_{mn}x_L^mx_L^n$, $\chi^{\alpha\dot{\beta}}_{mn}x_L^mx_R^n$ and $\sigma^{\alpha\dot{\beta}}_{mn}x_R^mx_R^n$ satisfying both (\ref{LeftDiracEqn}) and 
(\ref{RightDiracEqn}). Looking at the part linear in $x_R$, this implies:
\begin{align}
   \left(\Gamma_m \hat{x}_R\hat{F}\right)^{\alpha\dot{\beta}}  \;=\;
   - \;2 \sigma^{\alpha\dot{\beta}}_{mn}x_R^n
+  \chi_{mn}^{\alpha\dot{\beta}}x_R^n 
\end{align}
The left Dirac equation on $\chi$ implies $\Gamma^m_{\alpha\alpha'}\chi^{\alpha'\dot{\beta}}_{mn} = 0$, therefore:
\begin{equation}
   10\left(\hat{x}_R\hat{F}\right)_{\alpha}^{\dot{\beta}} = -2 \Gamma^m_{\alpha\alpha'}\sigma_{mn}^{\alpha'\dot{\beta}}x_R^n
\end{equation}
This implies that $\sigma$ is of the form:
\begin{align}
   \sigma^{\alpha\dot{\beta}}_{mn} =\;& - 5\delta_{mn}\hat{F}^{\alpha\dot{\beta}} + 
s_{mn}^{\alpha\dot{\beta}}
\label{DefS}
\\    
\mbox{ \tt where } \; & \Gamma^m_{\alpha\alpha'}s^{\alpha'\dot{\beta}}_{mn} = 0
\label{LeftDiracOnS}
\end{align}
for some $s_{mn}^{\alpha\dot{\beta}}$ symmetric in $m\leftrightarrow n$. As we have already mentioned, $\sigma$ should 
satisfy the right Dirac equation: 
\begin{equation}\label{RightDiracOnSigma}
   \sigma_{mn}^{\alpha\dot{\beta}'}\Gamma^n_{\dot{\beta}'\dot{\beta}} = 0
\end{equation}
Equations (\ref{LeftDiracOnS}) and (\ref{RightDiracOnSigma}) imply that the traces of $\sigma$ and $s$ are zero:
\begin{equation}
   \sigma_{mm}^{\alpha\dot{\beta}} = s_{mm}^{\alpha\dot{\beta}} = 0
\end{equation}
but this contradicts (\ref{DefS}) because the trace of $\delta_{mn} \hat{F}$ is not zero. This shows
that  (\ref{Cocycle}) is not in the image of $c^m\left(
   {\partial\over\partial x_L^m} - {\partial\over\partial x_R^m}
\right)$, and therefore it 
represents a nonzero element of the cohomology group (\ref{GroupH1}). This implies that
$\Phi[F]$ is a BRST-nontrivial vertex operator at the ghost number three. 

\paragraph     {Generalization} 
The cohomology of $Q_{{\rm SMaxw}\otimes {\rm SMaxw}}$ at the ghost number 3 is trivial, {\it i.e.}
any cocycle with three $\lambda$'s can be represented as $Q_{{\rm SMaxw}\otimes {\rm SMaxw}}\Psi$. But 
sometimes $\Psi$ cannot be chosen to depend on $x_L$ and $x_R$ through $x_L + x_R$ only.
The obstacle for that is in $H^1({\bf R}^{10},\;{{\rm SMaxw}\otimes {\rm SMaxw}})$ where ${\bf R}^{10}$ is the 
abelian group of translations, the Lie cohomology differential is 
$Q_{\rm Lie} = c^m\left({\partial\over\partial x_L^m} - {\partial\over\partial x_R^m}\right)$. Notice that ${{\rm SMaxw}\otimes {\rm SMaxw}}$ splits into 
components:
\begin{align}
   & {{\rm SMaxw}\otimes {\rm SMaxw}} = 
\\    
= \;&({{\rm Maxw}\otimes {\rm Maxw}})
\oplus 
({{\rm Maxw}\otimes {\rm Dirac}})
\oplus 
({{\rm Dirac}\otimes {\rm Maxw}})
\oplus 
({{\rm Dirac}\otimes {\rm Dirac}})
\nonumber
\end{align}
Consider the cohomology in the  sector ${{\rm Dirac}\otimes {\rm Dirac}}$, and more specifically
those elements of it which have linear $x$-dependence. It turns out that this
cohomology is identified with the quadratic in $x$ solutions $f$ of the 
``double Dirac equation'' modulo solutions presentable as a sum of a solution
of the left Dirac equation and a solution of the right Dirac equation:
\begin{align}
\;&
   {\partial\over\partial x^m}\Gamma^m_{\alpha\alpha'}
   {\partial\over\partial x^n}\Gamma^n_{\dot{\alpha}\dot{\alpha}'}
   f^{\alpha'\dot{\alpha}'}(x) = 0 
\nonumber \\[5pt]
\mbox{\tt but }\nexists \; s \mbox{ \tt and }\sigma \mbox{ \tt such that: }& 
f^{\alpha\dot{\alpha}} = s^{\alpha\dot{\alpha}} + \sigma^{\alpha\dot{\alpha}}
\label{DoNotExistSAndSigma}\\   
\;&   {\partial\over\partial x^m}\Gamma^m_{\alpha\alpha'}s^{\alpha'\dot{\alpha}} =0 
\;\mbox{ \tt and }
{\partial\over\partial x^n}
\sigma^{\alpha\dot{\alpha}'}\Gamma^n_{\dot{\alpha}'\dot{\alpha}} =0
\nonumber
\end{align}
Indeed, given such an $f^{\alpha\dot{\alpha}}$ with the quadratic $x$-dependence, we construct $\psi(c)$ 
in the following way:
\begin{equation}
   \psi(c) = \hat{c}\Gamma^n{\partial\over\partial x_R^n} f(x_R) + 
\xi(x_L,c)
\end{equation}
where $\xi$ is some solution of the left Dirac equation, chosen so that 
$Q_{\rm Lie}\psi = 0$; such a solution always exists  because $H^2({\bf R}^{10},\;{\rm Dirac}) = 0$. 
Suppose that $\psi$ is in the image of $Q_{\rm Lie}$ acting on the quadratic (in $x_{L|R}$) 
elements of ${\rm Dirac}\otimes {\rm Dirac}$, {\it i.e.}: 
\begin{equation}
   \psi(c) \stackrel{?}{=} c^m\left(
      {\partial\over\partial x_L^m} - {\partial\over\partial x_R^m}
   \right)\left(
      \sigma\langle x_R\otimes x_R\rangle +
      \chi\langle x_R\otimes x_L\rangle +
      \phi\langle x_L\otimes x_L\rangle
   \right)
\end{equation}
The part of $\psi(c)$ linear in $x_R$ would be:
\begin{equation}
- c^m{\partial\over\partial x_R^m}\sigma\langle x_R^{\otimes 2}\rangle
+ c^m{\partial\over\partial x_L^m}\chi\langle x_R\otimes x_L\rangle
\end{equation}
This implies:
\begin{equation}
   \Gamma^m{\partial\over\partial c^m}\psi(c)\langle x_R\rangle = 
10 \Gamma^n{\partial\over\partial x_R^n} f(x_R) = - \Gamma^m{\partial\over\partial x_R^m}\sigma\langle x_R^{\otimes 2}\rangle
\end{equation}
in other words $f = s + \sigma$ where $\sigma$ satisfies the right Dirac equation and $s$ 
the left Dirac equation. This contradicts (\ref{DoNotExistSAndSigma}).

Eq. (\ref{DefS}) has $f^{\alpha\dot{\alpha}} = ||x||^2 \hat{F}^{\alpha\dot{\alpha}}$ with a 5-form $\hat{F}$; there are also solutions 
corresonding to a 3-form or 7-form $\hat{G}$: 
\begin{equation}
   f = \hat{G} ||x||^2 - 
   {1\over 52} \hat{x}\Gamma_p\hat{G}\Gamma^p\hat{x}
\end{equation}
and a 1-form or 9-form $\hat{A}$:
\begin{equation}
   f = \hat{A} ||x||^2 - 
   {1\over 28} \hat{x}\Gamma_p\hat{A}\Gamma^p\hat{x}
\end{equation}
This means that the cohomology at the ghost number 3 at least includes
states with the quantum number of a bispinor. 

\subsubsection{Dual picture}
We conjecture that the dual element of $H_3(I)$ is of the form:
\begin{align}
   {\cal O}^{\alpha\dot{\beta}} = \;&\phantom{-}
   [\nabla_m^L,W_L^{\alpha}] \wedge W_R^{\dot{\beta}} \wedge 
   (\nabla^L_m - \nabla^R_m)\;-
\nonumber \\   
\;& - W_L^{\alpha}\wedge [\nabla_m^R,W_R^{\dot{\beta}}] \wedge 
   (\nabla^L_m - \nabla^R_m)\;+
\nonumber \\  
\;& +{1\over 2} 
W_L^{\alpha}\wedge W_R^{\dot{\beta}'} (\Gamma^{mn})^{\dot{\beta}}_{\dot{\beta}'}
\wedge [\nabla^R_m,\nabla^R_n]\; +
\nonumber \\   
\;& +{1\over 2}
 W_L^{\alpha'}(\Gamma^{mn})^{\alpha}_{\alpha'}\wedge W_R^{\dot{\beta}} 
\wedge [\nabla^L_m,\nabla^L_n]
\end{align}

\subsubsection{Conjecture about the vertices at the ghost number 3}
Generally speaking, the physical interpretation of vertex operators is:
\begin{itemize}
   \item Ghost number 1: global symmetries of the space-time
   \item Ghost number 2: infinitesimal deformations of the space-time
   \item Ghost number 3: obstructions to continuing the infinitesimal
      deformations of the space-time to the second order in the deformation
      parameter
\end{itemize}
It is natural to conjecture that the vertices at the ghost number 3 obstruct 
those and only those infinitesimal deformations which are unphysical in the 
sense of \cite{Mikhailov:2012id}.

\vspace{10pt}

\noindent
The cohomology at the ghost numbers 3 and 4 deserves systematic investigation.
We hope to return to this subject in the future work.

\section{Conclusion}
In this paper we presented a relation between the cohomology of the pure spinor
BRST complex in AdS space and the relative Lie algebra cohomology.

We used this relation to develop a ``dual'' point of view on the vertex 
operators in Type IIB. In this approach, instead of looking at the vertex 
operators, we look at the dual linear space which is identified with the 
gauge-invariant local operators of the Type IIB SUGRA. This works both in 
flat space and in AdS. We observe that some elements of the BRST cohomology
do not correspond to any physical states, {\it e.g.} the $A^+$ of (\ref{DefApm}). 
It turns out that there are also vertex operators at the ghost number three. 
They correspond to the obstructions for nonlinear deformations in the actions. 
Physically, these obstructions should not be present. 

Such ``unphysical'' elements should go away if we restrict the BRST complex to 
the operators annihilated by the Virasoro constraints. We do not know what 
this restriction means from the point of view of the Lie algebra cohomology. 

We conclude that the BRST complex (\ref{StandardBRSTComplex}) in $AdS_5\times S^5$ and its flat space 
limit (\ref{Qflat}) both have rich mathematical structure. But at the same time the 
cohomology does not give a complete description of the supergravity 
excitations. The difference is in some unphysical states. These unphysical
states have polynomial $x$-dependence, as opposed to the usually considered 
exponential $x$-dependence.  This polynomial (or ``zero-momentum'') sector 
could be important in the calculation of the scattering amplitude, because the
momentum conservation implies that the product of the scattering vertices 
has zero total momentum.

\appendix

\section{Exactness of (\ref{RelativeResolution})}\label{sec:Exactness}
This is similar to the proof of the exactness of the standard Koszul 
resolution of the Lie algebra in \cite{Knapp}. For any Lie algebra $L$, the universal
enveloping $UL$ is filtered so that ${\bf gr}^p UL = F^pUL/F^{p-1}UL = S^pL$. The
differential in our complex acts in such a way, that we can consistently
define:
\begin{align}
\ldots \longrightarrow 
F^{p-2}U{\cal L}^{\rm tot}\otimes_{{\bf g}_0} (\Lambda^2 I \otimes_{\bf C} A)
\longrightarrow 
F^{p-1}U{\cal L}^{\rm tot}\otimes_{{\bf g}_0} (I \otimes_{\bf C} A)
\longrightarrow \;&  
\nonumber \\    
\longrightarrow 
F^p U{\cal L}^{\rm tot}\otimes_{{\bf g}_{\bar{0}}} A
\longrightarrow 
F^p U{\bf g}\otimes_{{\bf g}_{\bar{0}}} A \longrightarrow \;& 0
\end{align}
This defines a series of complexes $d: X_n^p \to X_{n-1}^p$ parametrized by an 
integer $p$, where $X_{-1}^p = F^p U{\bf g}\otimes_{{\bf g}_{\bar{0}}} A$, $X_0^p=F^p U{\cal L}^{\rm tot}\otimes_{{\bf g}_{\bar{0}}} A$, and
$X_n^p = F^{p-n}U{\cal L}^{\rm tot}\otimes_{{\bf g}_0} (\Lambda^n I \otimes_{\bf C} A)$ for $n>0$. 
At $p=0$ we get the exact sequence:
\begin{equation}
   0\longrightarrow A \longrightarrow A \longrightarrow 0
\end{equation}
On the other hand, the factor-complex $X^p/X^{p-1}$ is:
\begin{align}
\ldots \longrightarrow 
S^{p-2}\left({\cal L}^{\rm tot}/{\bf g}_{\bar{0}}\right)
\otimes_{{\bf C}} \Lambda^2 I \otimes_{\bf C} A
\longrightarrow 
S^{p-1}\left({\cal L}^{\rm tot}/{\bf g}_{\bar{0}}\right)\otimes_{\bf C} 
I \otimes_{\bf C} A
\longrightarrow \;&  
\nonumber \\    
\longrightarrow 
S^p \left({\cal L}^{\rm tot}/{\bf g}_{\bar{0}}\right)\otimes_{\bf C} A
\longrightarrow 
S^p \left({\bf g}/{\bf g}_{\bar{0}}\right)
\otimes_{\bf C} A \longrightarrow \;& 0
\end{align}
This is exact, being the de Rham complex of the linear space $I$ times 
functions of additional ``inert'' variables corresponding to a complement to 
${\bf g}_{\bar{0}} + I$ in ${\cal L}^{\rm tot}$. By induction, the complexes $X^p$ are exact for all values 
of $p$, and therefore the complex (\ref{RelativeResolution}) is exact.

\section*{Acknowledgments}
I would like to thank Nathan Berkovits for discussions and the anonymous 
referee for useful suggestions.
This work was supported in part by the Ministry of Education and Science of 
the Russian Federation under the project 14.740.11.0347 ``Integrable and 
algebro-geometric structures in string theory and quantum field theory'', and 
in part by the RFFI grant 10-02-01315 
``String theory and integrable systems''.


\def\cprime{$'$} \def\cprime{$'$}
\providecommand{\bysame}{\leavevmode\hbox to3em{\hrulefill}\thinspace}
\providecommand{\MR}{\relax\ifhmode\unskip\space\fi MR }
\providecommand{\MRhref}[2]{%
  \href{http://www.ams.org/mathscinet-getitem?mr=#1}{#2}
}
\providecommand{\href}[2]{#2}

\end{document}